\documentclass[12pt]{article}
\usepackage{paper2e}
\usepackage{epsfig}

\begin{document}
\begin{titlepage}

\preprint{UTAP-528, RESCEU-9/05}

\title{Warped Flux Compactification and Brane Gravity}

\author{Shinji Mukohyama$^{\rm a,b}$, 
Yuuiti Sendouda$^{\rm a}$,\\
Hiroyuki Yoshiguchi$^{\rm a}$,
Shunichiro Kinoshita$^{\rm a}$
}

\address{$^{\rm a}$Department of Physics\\
The University of Tokyo, Tokyo 113-0033, Japan}

\address{$^{\rm b}$Research Center for the Early Universe\\
The University of Tokyo, Tokyo 113-0033, Japan}

\begin{abstract}
We find a simple exact solution of $6$-dimensional braneworld which
captures some essential features of warped flux compactification,
including a warped geometry, compactification, a magnetic flux, and one
or two $3$-brane(s). In this setup we analyze how the Hubble expansion
rate on each brane changes when the brane tension changes. It is shown
that effective Newton's constant resulting from this analysis agrees
with that inferred by simply integrating extra dimensions out. Based on 
the result, a general formula for effective Newton's constant is
conjectured and its application to cosmology with type IIB warped string
compactification is discussed. 
\end{abstract}

\end{titlepage}

\section{Introduction}

Warped flux compactification is one of essential parts of the 
construction of de Sitter vacua in string theory by Kachru, Kallosh,
Linde and Trivedi~\cite{KKLT}~\footnote{
See \cite{EGQ,BKQ} for followup proposals. See also
\cite{Silverstein,BCK}, 
\cite{Townsend-Wohlfarth,Ohta,Roy,Neupane} and \cite{BBK} for other
proposals of de Sitter, transiently accelerating and inflationary
universes.}. 
A warped throat region~\cite{KS} is smoothly attached to a Calabi-Yau
manifold and all moduli except for the volume modulus are stabilized by 
fluxes~\cite{GKP}. The volume modulus is thought to be stabilized by
non-perturbative effects such as
D-instantons~\cite{Witten,KKLT}. Another important ingredient of the
construction is branes. Anti-D-branes are located at the bottom of the 
throat region to uplift the stable AdS vacua to meta-stable de Sitter 
vacua. This setup provides a number of possible applications to
cosmology~\cite{KKLMMT,BBCEGKLQ,InflationKKLTsetup,Polchinski,Mukohyama2005}.

So far, there is no known explicit form of the global geometry including
the throat region, compactification, fluxes and the moduli
stabilization. The lack of known explicit form of the global geometry
makes it difficult to analyze brane gravity in the context of warped
flux compactification in details from higher dimensional point of
view. Of course, even without the explicit global geometry, one could
invoke that the $4$-dimensional Einstein gravity should be recovered as
a low energy effective theory since all moduli are thought to be
stabilized and made massive so that they can be integrated
out. Nonetheless, it is perhaps fair to say that we have not yet had a
complete understanding of how $4$-dimensional Einstein gravity is
recovered in the warped flux compactification.

For example, suppose that a D-brane and an anti-D-brane annihilate
somewhere in the warped throat region but that they are not coincident
with a brane on which we are living. Following the idea of
\cite{Dvali-Tye,Quevedo}, it is suggested that this process
may drive an inflation in our $4$-dimensional universe~\cite{KKLMMT} and
could leave cosmic superstrings as relics~\cite{Polchinski}. In this
picture the inflaton and cosmic superstrings are not on our brane but
living somewhere in the extra dimensions. It does not seem completely
clear how they affect gravity on our brane. Indeed, a lot of questions
would arise. ``Does the inflaton living outside our world really inflate
our brane?'' ``Is there a deficit angle on our brane due to the cosmic
superstring wandering somewhere in the extra dimensions?'' ``If there is
a deficit angle, then where on our brane and how much?'' Partial answers
to these questions would be obtained if we would completely integrate
out extra degrees of freedom due to the extra dimensions, provided that
all moduli are stabilized. On the other hand, in principle it should
also be possible to answer these questions by analyzing the higher
dimensional theory directly. Evidently, it is important to take both
approaches complementarily and compare them towards our good
understanding of brane gravity in the warped flux compactification.

Now let us heuristically remind ourselves how $4$-dimensional Einstein
gravity is recovered in the Kaluza-Klein compactification and
the Randall-Sundrum type braneworlds.

In Kaluza-Klein compactification, zero modes and Kaluza-Klein (KK) modes
are decoupled in the linearized level because of the momentum
conservation along extra dimensions. This simple fact makes the recovery
of the $4$-dimensional theory manifest since the standard model 
fields are supposed to consist of zero modes. Indeed, as far as moduli
associated with compactification are made massive by a stabilization
mechanism, they do not appear in the low energy physics and the  
Einstein theory is recovered as a low energy effective theory. This
consideration also applies to cosmology as far as the energy scale is
sufficiently lower than the compactification scale.

In the second Randall-Sundrum (RS2) braneworld with infinite extra
dimension~\cite{RS2}, the recovery of the $4$-dimensional Einstein
theory is due to localization of zero
mode~\cite{RS2,Garriga-Tanaka,SSM,GKR}. Matter on our brane possibly
excites not only zero modes but also KK modes since the momentum
conservation does not prohibit couplings between the singular brane
source and the KK modes~\footnote{
Note that a delta function includes all momenta when it is Fourier 
transformed.
}.
However, the warped geometry localizes the zero mode to the vicinity of
the brane and the coupling of the brane source to the zero mode is much
stronger than that to KK modes. In this way, gravity on our brane
at low energy is almost locally determined by the zero mode localized
near the brane and the $4$-dimensional Einstein theory is recovered at
low energy. (This point of ``locally localized gravity'' has 
been made explicit in ref.~\cite{Karch-Randall}.) For FRW cosmology on
the brane, there is no well-defined distinction between zero modes and
KK modes because of the lack of enough symmetry. Nonetheless, the
evolution of the brane is still determined locally since the unbroken
symmetry, i.e. the homogeneity and the isotropy parallel to the brane,
prevents waves from being generated and propagating in the bulk. In this
way, gravity is still localized and the standard cosmological equation
is recovered at low
energy~\cite{SMS,CGKT,CGS,FTW,BDEL,Mukohyama2000a,MSM,Kraus,Ida}.

On the other hand, in the warped compactification the recovery of
$4$-dimensional theory seems more indirect and subtle. Unlike KK
compactification but as in RS2 braneworld, matter source on the brane
can excite not only zero modes but also KK modes. However, unlike the 
RS2 braneworld, gravity is not localized near the brane since the warp
factor on the brane in the throat region is not larger but smaller than
that in the bulk nearby. Hence, the evolution of matter on the brane
changes the bulk geometry not only near the brane but possibly
everywhere in the whole extra dimensions. Nonetheless, if all moduli are
stabilized, the bulk geometry should quickly settle to a configuration
which is determined by the boundary condition, i.e. the brane
source(s), values of conserved quantities and the regularity of the
other region of the extra dimensions. As a consequence of the change of
the bulk geometry, the induced geometry on the brane responses to the
evolution of the matter source on the brane. It is quite possible but
must be checked that the $4$-dimensional Einstein theory is recovered as
a rather indirect and subtle relation between the matter source on the
brane and the response of the induced geometry. In this paper we shall
support this picture in a simplified situation.

This picture is somehow similar to that in the first Randall-Sundrum
(RS1) scenario~\cite{RS1} with radion stabilization~\cite{GW}. In the
original RS1 brane model (without radion stabilization) the
$4$-dimensional theory is not Einstein but Brans-Dicke
theory~\cite{Garriga-Tanaka,SSM} because of the existence of a massless 
modulus called radion. With the radion stabilization, there is no
extra massless degree which could appear in the $4$-dimensional
effective theory and the $4$-dimensional Einstein theory is recovered at
energies sufficiently below the stabilization scale. The recovery of the
$4$-dimensional Einstein theory is understood both from the low energy 
effective theory point of view and as a consequence of dynamics of bulk 
fields~\cite{Tanaka-Montes,CGK,Mukohyama-Kofman,Mukohyama-HDterm,Kudoh-Tanaka,CGR,KKOP}.

In the case of KK compactification and RS braneworlds, the recovery of
the Einstein theory has been explicitly investigated from higher
dimensional point of view. One of the reasons why this was possible is
that there are explicit background solutions (or at least explicit
equations defining background solutions) around which we can analyze
perturbations. On the other hand, in the warped flux compactification,
the absence of an explicit global solution makes it less tractable to
see the recovery of the $4$-dimensional Einstein gravity from higher
dimensional point of view as explicitly as in the KK compactification
and the RS braneworlds.

Therefore, we would like to consider a simplified situation in which we
can see the recovery of the $4$-dimensional Einstein theory in the
warped flux compactification. The purpose of this paper is, as a first
step, to consider a toy model which captures some essential features of
the warped flux compactification and to see explicitly how the induced
geometry on a brane responses to brane tension as a consequence of
changes in the bulk geometry. In particular, we shall see that the
relation between the change of brane tension and the response of the
induced geometry is identical to that inferred from the $4$-dimensional
Einstein theory.

The rest of this paper is organized as follows. In Sec.~\ref{sec:model}
we describes a simple model of $6$-dimensional warped flux
compactification with one or two $3$-brane(s). In
Sec.~\ref{sec:Friedmann-eq} we argue that the $4$-dimensional Friedmann
equation should be recovered on each brane at low energy and confirms
the validity of a formula of effective Newton's constant in a
simplified situation. Sec.~\ref{sec:summary} is devoted to a summary of
this paper and discussion. In particular, based on the result of
this paper, a general formula for effective Newton's constant is
conjectured and its application to cosmology with type IIB warped string
compactification is discussed. In Appendices~\ref{app:alphato1} and
\ref{app:Hto0} we show that two seemingly singular limits of the
solutions considered in this paper are actually regular. In
Appendix~\ref{app:rho2-correction} a higher-order correction to the
effective Friedmann equation on a brane is estimated and it is shown 
that higher order corrections can be ignored when the Hubble expansion
rate on the brane is sufficiently lower than the bulk curvature scale.

\section{Model description}
\label{sec:model}

In this paper we would like to consider a minimal setup which captures
essential features of the warped flux compactification and which is
simple enough to analyze brane gravity from higher dimensional
viewpoint. The setup must include at least a warped geometry, magnetic
flux of an antisymmetric field along the extra dimensions and a
brane. Since the simplest antisymmetric field is a $U(1)$ gauge field
and the corresponding magnetic flux has two spatial indices, we need to
consider at least $2$ extra dimensions and, thus, at least
$6$-dimensional spacetime. In this section we shall describe a model of
$6$-dimensional braneworld with warped flux compactification. In the
next section we shall analyze this model in a situation where the
tension of a $3$-brane changes by a phase transition on the brane.

For the reason explained above, in this paper we consider a
$6$-dimensional braneworld scenario with a $U(1)$ gauge field and a
cosmological constant in the bulk. The bulk action is 
%
\begin{equation}
 I_6 = 
  \frac{M_6^4}{2}\int d^6x\sqrt{-g}
  \left(R-2\Lambda_6-\frac{1}{2}F^{MN}F_{MN}\right),
\end{equation}
where $M_6$ is the $6$-dimensional reduced Planck mass, $\Lambda_6$ is
the bulk cosmological constant, and $F_{MN}=\partial_MA_N-\partial_NA_M$
is the field strength associated with the $U(1)$ gauge field $A_M$.

\subsection{Bulk solution}
\label{subsec:bulk}

We assume that the $6$-dimensional bulk geometry has the $4$-dimensional
de Sitter symmetry and an additional axisymmetry. The former symmetry is
to make the induced geometry on our brane a de Sitter spacetime, and the
latter is imposed as a symmetry in the extra dimensions for
simplicity. The general metric with these symmetries is locally written
as 
%
\begin{equation}
 ds_6^2 = A(w)^2ds_{4}^2 + dw^2 + R(w)^2d\phi^2,
  \label{eqn:6d-metric}
\end{equation}
where $w$ is the proper distance along geodesics orthogonal to the
brane world-volume and to the orbit of the axisymmetry, $A(w)$ and
$R(w)$ are functions of $w$ only, and $ds_4^2$ is the line element of
the $4$-dimensional de Sitter spacetime with the unit curvature
radius. For example, in the global chart $ds_4^2$ is written as 
%
\begin{equation}
 ds_4^2 = -dt^2+\cosh^2td\Omega_3^2,
\end{equation}
where $d\Omega_3^2$ is the line element of the unit $3$-sphere.

If $A(w)$ is constant then the solution to the Einstein equation 
is of the ADD type~\cite{ADD}. This solution was already investigated in
\cite{Carroll-Guica,Garriga-Porrati,Navarro}. In
Appendix~\ref{app:alphato1}, we shall see that this solution is a 
particular limit of the solutions below.

Our interest is in the case of non-vanishing $\partial_wA$ since this
corresponds to a warped geometry~\footnote{
This setup was considered in \cite{Vinet-Cline,CDGV} but, as far as we
know, the family of explicit exact solutions of warped flux
compactification presented below had not been found in the
literature.}. (As already stated, we would like to consider a minimal 
setup which includes at least a warped geometry, a flux and a brane.) In
this case we can introduce a new coordinate $r$ by $r=A(w)$ at least 
locally. With the new coordinate the line element is 
%
\begin{equation}
 ds_6^2 = r^2ds_4^2 + g(r)dr^2 + f(r)d\phi^2,
  \label{eqn:6d-metric-mod}
\end{equation}
where $f(r)$ and $g(r)$ are functions of $r$ only. It is of course
possible and indeed straightforward to analyze the Einstein equation and
find solutions with this ansatz. It is also possible to take double Wick
rotation, find a family of solutions and then take inverse double Wick
rotation. By the double Wick rotation 
%
\begin{equation}
 t \to i\left(\frac{\pi}{2}-\theta\right), \quad \phi \to iT,
\end{equation}
the metric ansatz (\ref{eqn:6d-metric-mod}) is transformed to 
%
\begin{equation}
 d\tilde{s}_6^2 = -\tilde{f}(r)dT^2 + \tilde{g}(r)dr^2 + r^2d\Omega_4^2, 
  \label{eqn:metric-sph}
\end{equation}
where 
%
\begin{equation}
 d\Omega_4^2 = d\theta^2+\sin^2\theta d\Omega_3^2
\end{equation}
is the line element of the unit $4$-sphere. The metric
(\ref{eqn:metric-sph}) is nothing but a general ansatz for a spherically
symmetric, static metric. There is a well-known family of solutions:
RN-de Sitter (for $\Lambda_6>0$), RN (for $\Lambda_6=0$) and RN-AdS (for
$\Lambda_6<0$) spacetime. The RN-de Sitter, RN or RN-AdS solution is 
%
\begin{eqnarray}
 \tilde{f}(r) & = & \frac{1}{\tilde{g}(r)} = 
  1-\frac{\Lambda_6}{10}r^2-\frac{\mu}{r^3} + \frac{e^2}{12r^6}, \nonumber\\
 A_Mdx^M & = & \frac{e}{3r^3}dT,
\end{eqnarray}
where $\mu$ is a constant of integration corresponding to the mass
parameter and $e$ is the electric charge. Going back to the original
ansatz by the inverse double Wick rotation
%
\begin{equation}
 \frac{\pi}{2}-\theta \to  -it, \quad T \to -i\phi, 
\end{equation}
we obtain the family of solutions
%
\begin{eqnarray}
 f(r) & = & \frac{1}{g(r)} = 
  1-\frac{\Lambda_6}{10}r^2-\frac{\mu}{r^3} - \frac{b^2}{12r^6}, \nonumber\\
 A_Mdx^M & = & \frac{b}{3r^3}d\phi,
  \label{eqn:solution}
\end{eqnarray}
where we have introduced the magnetic charge $b$ by $e\to ib$ so that
$A_Mdx^M$ remains real.

\subsection{Brane sources}
\label{subsec:brane-sources}

In the following we shall consider one or two $3$-brane sources. For
this purpose we shall use the well-known formula
%
\begin{equation}
 \delta_{\pm} = \frac{\sigma_{\pm}}{M_6^4},
  \label{eqn:deficit-angle}
\end{equation}
where $\delta_{\pm}$ is the deficit angle due to the tension
$\sigma_{\pm}$ of the branes. We, however, have to keep in mind that
there is no simple general prescription, analogous to Israel's junction 
condition~\cite{Israel}, for obtaining physical characteristics of an
arbitrary distributional source with more than one
codimensions~\cite{Israel1975,Geroch-Traschen}. The difficulty is 
essentially due to the fact that Einstein equation is non-linear. The
formula (\ref{eqn:deficit-angle}) is valid under the axisymmetry if
radial stress is much smaller than energy density~\cite{FIU}. Namely,
with these conditions, $\sigma_{\pm}$ in (\ref{eqn:deficit-angle}) can
be effectively considered as inertial mass per brane's volume. On the
other hand, if radial stress is not negligible then the  formula
(\ref{eqn:deficit-angle}) should be considered as the definition 
of $\sigma_{\pm}$, which is in general different from inertial mass per
brane's volume.

Hereafter we assume that the function $f(r)$ given by
(\ref{eqn:solution}) has two positive roots $r=r_{\pm}$ ($0<r_-<r_+$)
and is positive between them ($r_-<r<r_+$). This requires that the
$6$-dimensional cosmological constant $\Lambda_6$ be positive. Thus,
hereafter we assume that $\Lambda_6>0$. Since $f$ vanishes at
$r=r_{\pm}$, the equation $r=r_{\pm}$ defines surfaces of codimension
$2$.

Near $r=r_{\pm}$ in the bulk ($r_-<r<r_+$), 
%
\begin{equation}
 \frac{dr^2}{f(r)} + f(r)d\phi^2 \simeq 
  d\rho_{\pm}^2 + \kappa_{\pm}^2\rho_{\pm}^2d\phi^2,
\end{equation}
where 
%
\begin{equation}
  \rho_{\pm} \equiv \sqrt{\frac{\mp 2(r-r_{\pm})}{\kappa_{\pm}}}, \quad
   \kappa_{\pm} \equiv \mp \frac{1}{2}f'(r_{\pm}) \ (>0).
   \label{eqn:def-rho-kappa}
\end{equation}
Hence, with the deficit angles $\delta_{\pm}$ at $r=r_{\pm}$,
respectively, $\phi$ is identified as
%
\begin{equation}
 \kappa_{\pm}\phi \sim \kappa_{\pm}\phi + (2\pi -\delta_{\pm}). 
\end{equation}
The identification at $r=r_+$ is consistent with that at $r=r_-$ if and
only if 
%
\begin{equation}
 \frac{2\pi -\delta_+}{2\pi -\delta_-} = \frac{\kappa_+}{\kappa_-}. 
  \label{eqn:ratio-kappa}
\end{equation}
This can be considered as a boundary condition since the l.h.s. is
specified by the brane sources and the r.h.s. can be written in terms of
the bulk parameters $\mu$ and $b$.

In this way, we can put $3$-branes at $r=r_{\pm}$ and consider the
region $r_-<r<r_+$ as the bulk spacetime. It is also possible to
consider a solution with only one $3$-brane at either $r=r_+$ or $r=r_-$
by setting $\delta_-=0$ or $\delta_+=0$. The Hubble parameter $H_{\pm}$
on the brane at $r=r_{\pm}$ is given by 
%
\begin{equation}
 H_{\pm} = \frac{1}{r_{\pm}}. 
\end{equation}

\section{Recovery of the Friedmann equation}
\label{sec:Friedmann-eq}

Let us consider a $(4+n)$-dimensional, general warped compactification 
%
\begin{equation}
 ds_{4+n}^2 = r^2 g^{(4)}_{\mu\nu}dx^{\mu}dx^{\nu}
  + \gamma_{ij}dy^idy^j,
  \label{eqn:warped-metric-general}
\end{equation}
where the $4$-dimensional metric $g^{(4)}_{\mu\nu}$ depends on the 
$4$-dimensional coordinates $x^{\mu}$ only, and the $n$-dimensional
metric $\gamma_{ij}$ and the warp factor $r$ depend only on the
coordinates $y^i$ of the compact extra dimensions. With this warped 
metric, the $(4+n)$-dimensional Einstein-Hilbert action includes the 
$4$-dimensional Einstein term:
%
\begin{equation}
 (M_{4+n})^{2+n}\int d^4xd^ny \sqrt{-g^{(4+n)}}R^{(4+n)} =
   (M_{4+n})^{2+n}\int d^ny \sqrt{\gamma}\ r^2
   \times \int d^4x\sqrt{-g^{(4)}}R^{(4)} + \cdots,
\end{equation}
where dots represent terms including derivatives of $r$, the curvature
of $\gamma_{ij}$ and so on. Therefore, if all moduli associated with the
extra dimensions are stabilized and if their masses are large enough
then we expect that the $4$-dimensional Einstein theory is recovered at
low energy and that Newton's constant $G_N$ on a brane at
$y^i=y_0^i$ should be given by 
%
\begin{equation}
 \frac{1}{8\pi G_N} = 
  (M_{4+n})^{2+n}\int d^ny 
  \sqrt{\gamma}\left[\frac{r(y)}{r(y_0)}\right]^2,
  \label{eqn:GN-formula}
\end{equation}
where the normalization factor $[r(y_0)]^{-2}$ has been included in
order to take into account the fact that the induced metric on the brane
is not $g^{(4)}_{\mu\nu}dx^{\mu}dx^{\nu}|_{y=y_0}$ but
$r^2g^{(4)}_{\mu\nu}dx^{\mu}dx^{\nu}|_{y=y_0}$.

This expectation is known to be correct for codimension $1$ ($n=1$)
braneworlds with radion 
stabilization~\cite{Tanaka-Montes,CGK,Mukohyama-Kofman,Mukohyama-HDterm,Kudoh-Tanaka,CGR,KKOP}~\footnote{
With the $Z_2$ symmetry, the integration over the bulk must be
multiplied by $2$ to take into account the fact that there are two
copies of the same bulk geometry.
}.

We expect that the formula (\ref{eqn:GN-formula}) should be correct also
for codimension $2$ or higher ($n\geq 2$) braneworlds if all moduli are
stabilized. In this paper, for simplicity, we consider the
$6$-dimensional braneworld described in the previous section and a
situation where the tension of one of the $3$-branes at $r=r_{\pm}$ 
changes by a phase transition on the brane. We suppose that the tension
is almost constant in deep inside the old and new phases. With this
setup, it is expected that the $4$-dimensional geometries on the 
brane deep inside the two phases are approximated by de Sitter
spacetimes with different Hubble expansion rates. What we should see is
the relation between the difference of tension in the two phases and the
corresponding change of the Hubble expansion rate. Indeed, we shall 
see below that at low energy, the relation is identical to that inferred
from the standard Friedmann equation, provided that $4$-dimensional
Newton's constant is given by the formula (\ref{eqn:GN-formula}). We 
shall also see what ``low energy'' exactly means.

By the phase transition, the bulk geometry should change since the
boundary condition set by the brane tension changes. In particular, the
bulk parameters $\mu$ and $b$ after the phase transition will in general
be different from what they were before. Actually, by the following
reason, the values of $\mu$ and $b$ after the phase transition should
be uniquely determined by the brane tension. Indeed, the boundary
condition (\ref{eqn:ratio-kappa}) and the conservation of magnetic flux
provide two independent conditions on the two independent parameters
$\mu$ and $b$ and, thus, uniquely fix the parameters at least for a
sufficiently small change.

When one of the brane tensions changes, the deficit angle at the
position of the brane changes, according to the formula
(\ref{eqn:deficit-angle}). This in general induces the change of the 
area of the extra dimensions, in particular the interval of
$\phi$. Since the magnetic flux is nothing but the integral of the
magnetic field over the extra dimensions, the change of the area and the
flux conservation imply that the amplitude $b$ of magnetic field should
change. At the same time, the parameters $\mu$ and $b$ must satisfy the
boundary condition (\ref{eqn:ratio-kappa}) with the new deficit angle
corresponding to the tension after the phase transition. Therefore, not
only $b$ but also $\mu$ should change in general. In this way, the
values of $b$ and $\mu$ after the phase transition are uniquely
determined by the flux conservation and the boundary condition.

Accordingly, positions $r_{\pm}$ of branes are also determined uniquely
since they are defined as roots of the function $f(r)$ in
(\ref{eqn:solution}). In other words, the Hubble expansion rate
$H_{\pm}=1/r_{\pm}$ changes after the phase transition and there is a 
unique relation between the change of the brane tension and the change
of the Hubble expansion rate. Since all relevant equations are invariant
under the reflection $H_{\pm}\to -H_{\pm}$, the resulting physical
relation must also be invariant under this reflection. Therefore, the
relation must be even in $H_{\pm}$ and its Taylor expansion
w.r.t. $\Delta\sigma_{\pm}$ (or equivalently, w.r.t.  
$\Delta H_{\pm}^2$) should start as 
%
\begin{equation}
 \Delta H_{\pm}^2 \propto \Delta \sigma_{\pm}
  + O((\Delta\sigma_{\pm})^2),
\end{equation}
where $\Delta X$ represents a small difference of the quantity $X$
before and after the phase transition. Finally, from this relation we
obtain 
%
\begin{equation}
 H_{\pm}^2 \propto (\sigma_{\pm} - \sigma_{\pm}^{(0)})
 + O((\sigma_{\pm}-\sigma_{\pm}^{(0)})^2) 
 \label{eqn:lowEH2}
\end{equation}
for small $H_{\pm}^2$, where $\sigma_{\pm}^{(0)}$ is the value of
$\sigma_{\pm}$ corresponding to $H_{\pm}^2=0$. In this argument, it has 
been implicitly assumed that the limit $H_{\pm}^2\to 0 $ is regular. 
In Appendix~\ref{app:Hto0} we explicitly confirm that this limit is
indeed regular.

The relation (\ref{eqn:lowEH2}) is nothing but the standard Friedmann
equation if the proportionality coefficient is $8\pi G_N/3$, where $G_N$
is Newton's constant. In the following, we shall see both numerically
and analytically that this is indeed the case, where Newton's
constant is given by the formula (\ref{eqn:GN-formula}). To be more
specific, we shall show that Newton's constant $G_{N\pm}$ on the
brane at $r=r_{\pm}$, respectively, is given by 
%
\begin{equation}
 \frac{1}{8\pi G_{N\pm}} = 
  M_6^4\int_{r_-}^{r_+}\left(\frac{r^2}{r_{\pm}^2}\right)
  dr\int_0^{\Delta\phi}d\phi 
  =  M_6^4L^2\Sigma\cdot\left(\frac{r_-}{r_+}\right)^{\pm 1},
  \label{eqn:NewtonConstBrane}
\end{equation}
where $L\equiv\sqrt{10/\Lambda_6}$, 
%
\begin{equation}
 \Sigma \equiv 
  \frac{1}{L^2}\int_{r_-}^{r_+}\frac{r^2}{r_+r_-}
  dr\int_0^{\Delta\phi}d\phi 
  = \frac{\Delta\phi}{3L^2r_+r_-}(r_+^3-r_-^3)
  \label{eqn:warped-volume}
\end{equation}
is the warped volume of the extra dimensions (in the unit of $L^2$), and
%
\begin{equation}
 \Delta\phi = \frac{2\pi-\delta_+}{\kappa_+}
  = \frac{2\pi-\delta_-}{\kappa_-}
\end{equation}
is the period of the angular coordinate $\phi$.

For later convenience, here we define the magnetic flux $\Phi$ in the
unit of $L$ as 
%
\begin{equation}
 \Phi = \frac{1}{L}\int_{r_-}^{r_+}dr\int_0^{\Delta\phi} d\phi F_{r\phi}
  = -\frac{b\Delta\phi}{3L}\left(\frac{1}{r_-^3}-\frac{1}{r_+^3}\right).
  \label{eqn:flux}
\end{equation}
All relevant equations including this are invariant under the reflection
$(b,\Phi)\to(-b,-\Phi)$. Thus, we do not need to keep track of the
sign of $\Phi$ as far as its sign relative to $b$ is correct. As already
stated, the magnetic flux is conserved and, thus, must be fixed during
the phase transition.

\subsection{Re-parameterization}

As already explained, we expect that the Friedmann equation should be
recovered at low energy as a consequence of the response of the bulk
geometry to the evolution of the brane source. In this respect it is not
physically relevant to keep track of the change of the bulk parameters
$\mu$ and $b$. What is physically important is, instead, the brane
tensions (or equivalently the deficit angles) and the Hubble expansion
rates on the branes (or equivalently the positions $r_{\pm}$ of the
branes).

Hence, it is useful to express the metric in terms of 
$h\equiv L\sqrt{H_+H_-}=L/\sqrt{r_+r_-}$ and 
$\alpha\equiv H_+/H_-=r_-/r_+$ instead of the original parameters $\mu$
and $b$. By the definition of $r_{\pm}$, the new parameter $\alpha$
satisfies $0<\alpha\leq 1$. (In Appendix~\ref{app:alphato1}, the
$\alpha\to 1$ limit is shown to be regular.) By solving $f(r_{\pm})=0$
w.r.t. $\mu$ and $b^2$, we obtain  
%
\begin{eqnarray}
 \frac{h^5\mu}{L^3} & = & -\frac{\beta_{7/2}}{\beta_1}
  + (\alpha^{3/2}+\alpha^{-3/2})h^2, \nonumber\\
 \frac{h^8 b^2}{L^6} & = & \frac{12\beta_2}{\beta_1} - 12h^2,
\end{eqnarray}
where $\beta_n\equiv \sum_{i=0}^{2n}\alpha^{i-n}$. 
With this parametrization, (\ref{eqn:ratio-kappa}) and (\ref{eqn:flux})
become 
%
\begin{equation}
 \frac{2\pi-\delta_+}{2\pi-\delta_-}
  = \frac{\gamma_+-3\beta_1^2h^2}{\gamma_--3\beta_1^2h^2}\alpha^4
  \label{eqn:ratio-angle}
\end{equation}
and 
%
\begin{equation}
 \frac{\Phi^2}{(2\pi-\delta_+)(2\pi-\delta_-)} 
  = \frac{16}{3}
  \frac{(\beta_2-\beta_1 h^2)\beta_1^3}
  {(\gamma_+-3\beta_1^2h^2)(\gamma_--3\beta_1^2h^2)},
  \label{eqn:flux-squared}
\end{equation}
respectively, where
%
\begin{eqnarray}
 \gamma_+ & = & 
  3\alpha^3+6\alpha^2+9\alpha+12+15\alpha^{-1}+10\alpha^{-2}+5\alpha^{-3},
  \nonumber\\
 \gamma_- & = &  
 5\alpha^3+10\alpha^2+15\alpha+12+9\alpha^{-1}+6\alpha^{-2}+3\alpha^{-3}.
\end{eqnarray}
By using $\Delta\phi>0$, $0<\alpha\leq 1$ and $b^2\geq 0$, it is shown
that 
%
\begin{equation}
 0 < \frac{2\pi-\delta_+}{2\pi-\delta_-} \leq 1,
  \label{eqn:eta-region}
\end{equation}
where the equality holds for $\alpha=1$. The inequality
(\ref{eqn:eta-region}) combined with the formula
(\ref{eqn:deficit-angle}) implies that the tension of the brane at 
$r=r_-$ must be smaller than the tension of the brane at $r=r_+$. If
this condition is not satisfied then it is expected that the
cosmological constant on the brane cannot be non-negative or/and the
geometry of the extra dimensions becomes dynamical. The warped volume
$\Sigma$ of extra dimensions defined in (\ref{eqn:warped-volume}) is
given by 
%
\begin{equation}
 \frac{\Sigma}{\sqrt{(2\pi-\delta_+)(2\pi-\delta_-)}} = 
  \frac{2\beta_1^2}
  {3\sqrt{(\gamma_+-3\beta_1^2 h^2)(\gamma_--3\beta_1^2 h^2)}}. 
  \label{eqn:WarpedVolume}
\end{equation}

\subsection{Brane gravity on the IR brane}
\label{subsec:IRbrane}

We now would like to see that the relation of the form
(\ref{eqn:lowEH2}) indeed holds at low energy on each brane. Since the
warp factor $r$ is smaller on the brane at $r=r_-$, we may call this
brane an IR brane and the other brane at $r=r_+$ a UV brane. In this and
the next subsections, we shall show the relation (\ref{eqn:lowEH2}) on
the IR and UV branes, respectively, by using numerical plots and
determine the proportionality coefficient. In
subsection~\ref{subsec:lowEexpansion} we show the same results analytically
by low energy expansion.

Let us first consider the case where the phase transition takes place on
the brane at $r=r_-$. In this case the deficit angle $\delta_+$
around the other brane at $r=r_+$ and the flux $\Phi$ must be fixed. It
is convenient to define the following dimensionless quantities:
%
\begin{eqnarray}
 \eta_- & \equiv & \frac{2\pi - \delta_-}{2\pi-\delta_+}, \nonumber\\
 h_- & \equiv & L^2H_- = h\alpha^{-1/2}, \nonumber\\
 \Phi_+ & \equiv & \frac{|\Phi|}{2\pi-\delta_+}
  = \left[\frac{\Phi^2}{(2\pi-\delta_+)(2\pi-\delta_-)}\right]^{1/2}
  \eta_-^{1/2}   \nonumber\\
 g_- & \equiv & 
   \frac{8\pi G_{N-}}{3}M_6^4L^2(2\pi-\delta_+)
   = \frac{1}{3}\left[\frac{\Sigma}{\sqrt{(2\pi-\delta_+)(2\pi-\delta_-)}}
     \right]^{-1}\eta_-^{-1/2}\alpha,
   \label{eqn:h-Phi-g-def}
\end{eqnarray}
where $G_{N-}$ is defined by (\ref{eqn:NewtonConstBrane}). The formulas
(\ref{eqn:ratio-angle}), (\ref{eqn:flux-squared}) and
(\ref{eqn:WarpedVolume}) give expressions of $\eta_-$, $\Phi_+^2$ and
$g_-$ in terms of $h_-^2$ and $\alpha$. The inequality
(\ref{eqn:eta-region}) is written as $\eta_-\geq 1$, where the equality
holds for $\alpha=1$.

The quantities $\alpha$, $h_-$, $\eta_-$, $\Phi_+$ and $g_-$ have the
following physical meaning. The first quantity $\alpha\equiv r_-/r_+$
($0<\alpha\leq 1$) is the ratio of the warp factor so that a small (or
large) value of $\alpha$ corresponds to a large (or small, respectively)
hierarchy. The second quantity $h_-$ is the Hubble expansion rate on the
brane at $r=r_-$ in the unit of the bulk curvature scale $L^{-1}$. Note
that in this subsection, the phase transition is supposed to take place
on this brane. Since the deficit angle $\delta_+$ around the other brane
at $r=r_+$ is determined by the tension of the other brane at $r=r_+$
(see the formula (\ref{eqn:deficit-angle})) and has nothing to do with
the phase transition on the brane at $r=r_-$, $2\pi-\delta_+$ is
constant during the phase transition. Hence, the change of tension of
the brane at $r=r_-$ is proportional to the change of the third quantity
$\eta_-$. The proportionality coefficient depends on $2\pi-\delta_+$ but
it does not change by the phase transition. Similarly, specifying the
magnetic flux up to its signature is equivalent to specifying the forth
quantity $\Phi_+$. As already stated, all relevant equations are
invariant under reflection of the sign of the flux and, thus, we do not
need to keep track of the sign. The normalization of $\Phi_+$ has been
determined so that it does not change during the phase transition and
that it absorbs the factor $(2\pi-\delta_+)^{-2}$ in the
eq.~(\ref{eqn:flux-squared}) divided by (\ref{eqn:ratio-angle}). The
final quantity $g_-$ is proportional to the quantity $G_{N-}$ which is
defined by (\ref{eqn:NewtonConstBrane}) and which shall be identified as
effective Newton's constant on the brane at $r=r_-$. The
normalization of $g_-$ has been determined so that $g_-$ is
dimensionless and that it eliminates explicit dependence on
$(2\pi-\delta_+)$ from eq.~(\ref{eqn:FriedmannEq-pre}) below.

When the Hubble expansion rate on the brane vanishes, i.e. when $h^2=0$, 
the equations are simplified. Hereafter, a superscript $(0)$ shows
that the corresponding quantity is evaluated at $h^2=0$. In particular,
it is easy to see that 
%
\begin{eqnarray}
 \frac{1}{5} &\leq& \Phi_+^{(0)2} < \infty,  \nonumber\\
 \frac{10}{3} &\geq& g_-^{(0)} > 0,
\end{eqnarray}
where each quantity (with $h=0$) is a monotonic function of $\alpha$ and
the left and right values are values at $\alpha=1$ and $\alpha=0$,
respectively.

\begin{figure}
 \centering\leavevmode\epsfysize=8cm \epsfbox{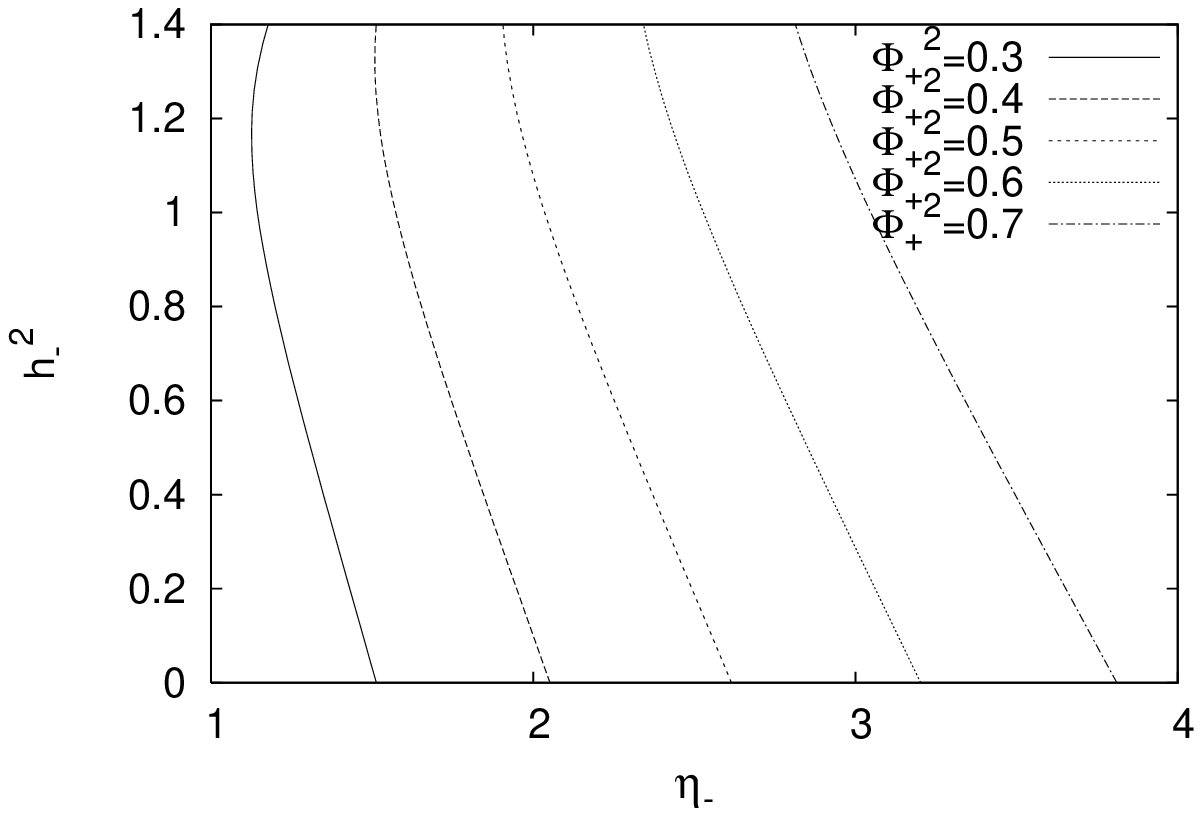}
 \caption{\label{fig:H2-eta}
 Each line represents the relation between $h_-^2$ (vertical axis) and
 $\eta_-$ (horizontal axis) for a fixed value of $\Phi_+^2$ ($=0.3$, 
 $0.4$, $0.5$, $0.6$, $0.7$ from left to right). Note that the value of
 $\eta_-$ is restricted to the region $\eta_-\geq 1$ as shown in
 (\ref{eqn:eta-region}). The domain shown in this figure can be joined
 to the domain shown in Fig.~\ref{fig:H2+eta} by identifying the
 vertical line $\eta_-=1$ in this figure to the vertical line $\eta_+=1$
 in Fig.~\ref{fig:H2+eta}.  
 }
\end{figure}

Our aim in this paper is to show that, when a phase transition takes
place on a brane, the Hubble expansion rate changes according to the
standard Friedmann equation with Newton's constant given by the
formula (\ref{eqn:NewtonConstBrane}). In the present context where the
phase transition takes place on the brane at $r=r_-$, we would like to
see a relation between $h_-^2$ and $\eta_-$ with $\Phi_+^2$
fixed. Actually, by using equations (\ref{eqn:ratio-angle}) and
(\ref{eqn:flux-squared}), it is fairly easy to plot the curve ($\eta_-$,
$h_-^2$) for various fixed values of $\Phi_+^2$. See
Fig.~\ref{fig:H2-eta}. It is seen that the slope of each curve in 
Fig.~\ref{fig:H2-eta} is negative near the horizontal axis, i.e. for
small $h_-^2$. In other words, at low energy the Hubble expansion rate
$h_-$ indeed decreases as the brane tension $\sigma_-$ and the deficit
angle $\delta_-$ decrease (see the formula (\ref{eqn:deficit-angle})) by
a phase transition, and Newton's constant on the brane is positive. The
intersection of each curve with the horizontal axis defines the value
$\eta_-^{(0)}(\Phi_+^2)$ of $\eta_-$ corresponding to $h_-^2=0$. Thus, 
%
\begin{equation}
 \sigma_-^{(0)}(\Phi_+^2) \equiv 
  \left[2\pi-(2\pi-\delta_+)\eta_-^{(0)}(\Phi_+^2)\right]\cdot M^4 
\end{equation}
is the critical value of tension $\sigma_-$ for which the effective
cosmological constant on the brane vanishes. If
$\sigma_->\sigma_-^{(0)}(\Phi_+^2)$ (or
$\sigma_-<\sigma_-^{(0)}(\Phi_+^2)$) then the effective cosmological
constant on the brane is positive (or negative, respectively). 

\begin{figure}
 \centering\leavevmode\epsfysize=8cm \epsfbox{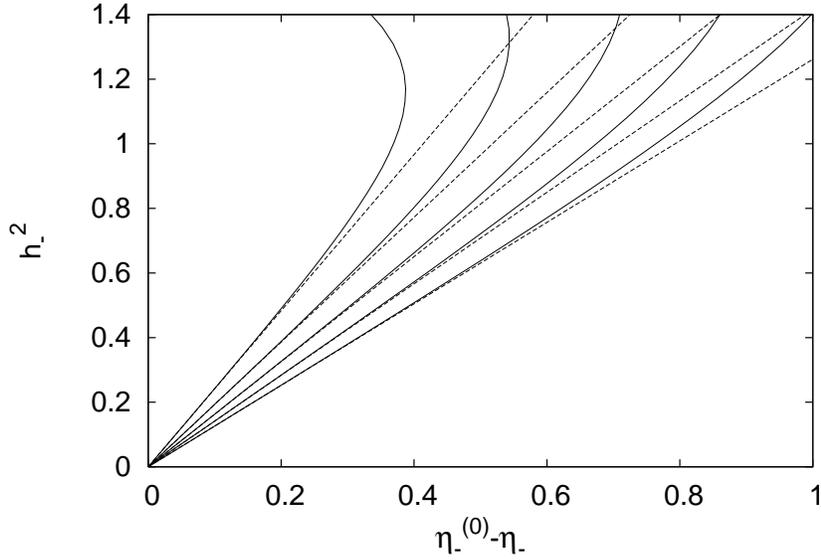}
 \caption{\label{fig:H-squared}
 Solid lines are curves shown in Fig.~\ref{fig:H2-eta} in a different
 coordinate. The vertical axis is again $h_-^2$ but the horizontal axis 
 is now $\eta_-^{(0)}(\Phi_+^2)-\eta_-$, where $\eta_-^{(0)}(\Phi_+^2)$
 is the value of $\eta_-$ corresponding to $h_-^2=0$ for a given value
 of $\Phi_+^2$. The value of $\Phi_+^2$ for each curve is $0.3$, $0.4$,
 $0.5$, $0.6$, $0.7$ from left to right. Dot lines represent straight
 lines passing through the origin with the slope $g_-^{(0)}(\Phi_+^2)$
 for each value of $\Phi_+^2$. These graphs confirm that $G_{N-}$
 defined by (\ref{eqn:NewtonConstBrane}) indeed has the meaning of 
 $4$-dimensional Newton's constant on the brane at $r=r_-$. 
 }
\end{figure}

The positivity of low energy effective Newton's constant is
highlighted in Fig.~\ref{fig:H-squared}, where the vertical axis is
again $h_-^2$ but the horizontal axis is now
$\eta_-^{(0)}(\Phi_+^2)-\eta_-$. The positive slope of the curve near
the origin clearly defines low energy effective Newton's constant
since it determines how much the Hubble expansion rate changes as the
brane tension changes. The dot lines in Fig.~\ref{fig:H-squared} are 
straight lines passing through the origin with the slope
$g_-^{(0)}(\Phi_+^2)$, where $g_-^{(0)}(\Phi_+^2)$ is the value of $g_-$
with $h_-^2=0$ for a given value of $\Phi_+^2$. It is easily seen that
the solid lines and the dot lines in Fig.~\ref{fig:H-squared} come in
contact with each other at the origin. This fact implies that 
%
\begin{equation}
 h_-^2 = g_-^{(0)}\cdot(\eta_-^{(0)}-\eta_-)
  + O((\eta_-^{(0)}-\eta_-)^2).
  \label{eqn:FriedmannEq-pre}
\end{equation}
By definition of $g_-$ and the formula (\ref{eqn:deficit-angle}), this
is equivalent to 
%
\begin{equation}
 H_-^2 = \frac{8\pi G_{N-}}{3}(\sigma_--\sigma_-^{(0)})
  + O((\sigma_--\sigma_-^{(0)})^2),
  \label{eqn:FriedmannEq-}
\end{equation}
which confirms that $G_{N-}$ defined in (\ref{eqn:NewtonConstBrane})
indeed has the meaning of $4$-dimensional Newton's constant on the
brane at $r=r_-$. The straight dot lines in Fig.~\ref{fig:H-squared}
are good approximation to the solid lines in the regime where $h_-^2$ is
sufficiently smaller than $1$. This suggests that the Friedmann equation
should be recovered at low energy where the Hubble expansion rate
squared $H_-^2$ on the brane is sufficiently smaller than the bulk
cosmological constant $\Lambda_{6}$ ($=10L^{-2}$). In
subsection~\ref{subsec:lowEexpansion}, we shall show the relation 
(\ref{eqn:FriedmannEq-}) analytically.  

\subsection{Brane gravity on the UV brane}
\label{subsec:UVbrane}

Next let us consider the case where a phase transition takes place on
the UV brane at $r=r_+$. In this subsection, we shall show the relation
(\ref{eqn:lowEH2}) on the UV brane by using numerical plots and
determine the proportionality coefficient. In the next
subsection~\ref{subsec:lowEexpansion} we show the same results
analytically by low energy expansion.

In the present case where a phase transition takes place on the brane at
$r=r_+$, the deficit angle $\delta_-$ around the brane at $r=r_-$ and
the flux $\Phi$ must be fixed. Similarly to the previous case, it is
convenient to define the following dimensionless quantities: 
%
\begin{eqnarray}
 \eta_+ & \equiv & \frac{2\pi - \delta_+}{2\pi-\delta_-}, \nonumber\\
 h_+ & \equiv & LH_- = h\alpha^{1/2}, \nonumber\\
 \Phi_- & \equiv & \frac{|\Phi|}{2\pi-\delta_-}
  = \left[\frac{\Phi^2}{(2\pi-\delta_+)(2\pi-\delta_-)}\right]^{1/2}
  \eta_+^{1/2}   \nonumber\\
 g_+ & \equiv & 
   \frac{8\pi G_{N+}}{3}M_6^4L^2(2\pi-\delta_-)
   = \frac{1}{3}\left[\frac{\Sigma}{\sqrt{(2\pi-\delta_+)(2\pi-\delta_-)}}
     \right]^{-1}\eta_+^{-1/2}\alpha^{-1},
\end{eqnarray}
where $G_{N+}$ is defined by (\ref{eqn:NewtonConstBrane}). The formulas
(\ref{eqn:ratio-angle}), (\ref{eqn:flux-squared}) and
(\ref{eqn:WarpedVolume}) give expressions of $\eta_+$, $\Phi_-^2$ and 
$g_+$ in terms of $h_+^2$ and $\alpha$. The inequality
(\ref{eqn:eta-region}) is written as $0<\eta_+\leq 1$, where the
equality holds for $\alpha=1$.

The quantities $h_+$, $\eta_+$, $\Phi_-$ and $g_+$ have the physical
meaning similar to $h_-$, $\eta_-$, $\Phi_+$ and $g_-$, whose physical
meaning has already been explained in the third paragraph of the
previous subsection.

When the Hubble expansion rate on the brane vanishes, i.e. when $h^2=0$, 
the equations are simplified. As before, a superscript $(0)$ shows
that the corresponding quantity is evaluated at $h^2=0$. In particular,
it is easy to see that 
%
\begin{eqnarray}
 0 & < & \Phi_-^{(0)2} \leq \frac{1}{5},  \nonumber\\
 \infty & > & g_+^{(0)} \geq \frac{10}{3}
\end{eqnarray}
where each quantities (with $h=0$) is a monotonic function of $\alpha$
and the left and right values are values at $\alpha=0$ and $\alpha=1$, 
respectively.

\begin{figure}
 \centering\leavevmode\epsfysize=8cm \epsfbox{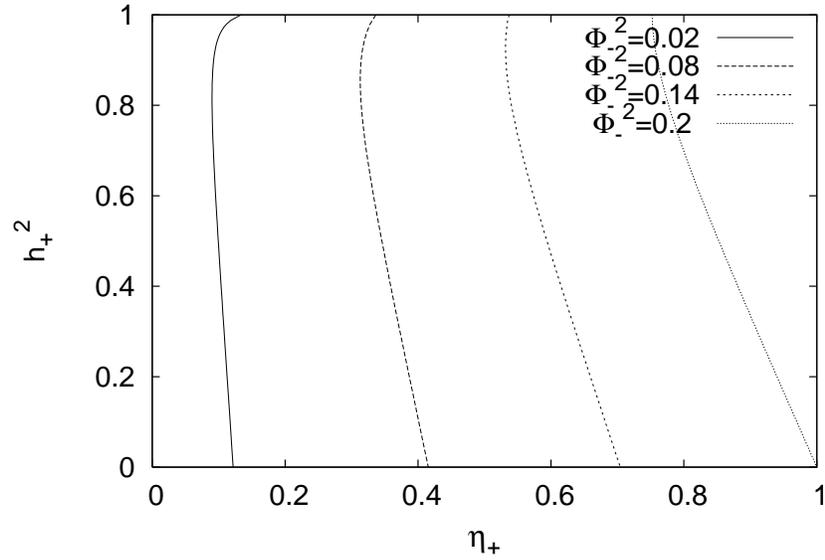}
 \caption{\label{fig:H2+eta}
 Each line represents the relation between $h_+^2$ (vertical axis) and
 $\eta_+$ (horizontal axis) for a fixed value of $\Phi_-^2$ ($=0.02$,
 $0.08$, $0.14$, $0.2$ from left to right). Note that the value of
 $\eta_+$ is restricted to the region $0<\eta_+\leq 1$ as shown in
 (\ref{eqn:eta-region}). For a relation between curves in this figure
 and Fig.~\ref{fig:H2-eta}, see the third-to-the-last paragraph in
 subsection~\ref{subsec:UVbrane}. The domain shown in this figure can be
 joined to the domain shown in Fig.~\ref{fig:H2-eta} by identifying the 
 vertical line $\eta_+=1$ in this figure to the vertical line $\eta_-=1$
 in Fig.~\ref{fig:H2-eta}.  
 }
\end{figure}

As already stated many times, our aim in this paper is to show that,
when a phase transition takes place on a brane, the Hubble expansion
rate changes according to the standard Friedmann equation with Newton's
constant given by the formula (\ref{eqn:NewtonConstBrane}). In this
subsection we are considering a situation where the phase transition
takes place on the brane at $r=r_+$. Thus, we would like to see a
relation between $h_+^2$ and $\eta_+$ with $\Phi_-^2$ fixed. Actually,
by using equations (\ref{eqn:ratio-angle}) and (\ref{eqn:flux-squared}),
it is fairly easy to plot the curve ($\eta_+$, $h_+^2$) for various
fixed values of $\Phi_-^2$. See Fig.~\ref{fig:H2+eta}. It is seen that
the slope of each curve in Fig.~\ref{fig:H2+eta} is negative near the
horizontal axis, i.e. for small $h_+^2$. In other words, at low energy
the Hubble expansion rate $h_+$ indeed decreases as the brane tension
$\sigma_+$ and the deficit angle $\delta_+$ decrease (see the formula
(\ref{eqn:deficit-angle})) by a phase transition, and Newton's constant
on the brane is positive. 

At first sight, this appears to contradict with the negativity of the
slope of the curves in Fig.~\ref{fig:H2-eta} near $h_-^2=0$:
Fig.~\ref{fig:H2+eta} indicates that $\eta_+$ decreases when $h^2$
increases from zero but Fig.~\ref{fig:H2-eta} indicates that
$\eta_-=1/\eta_+$ also decreases when $h^2$ increases from
zero. Actually, there is no contradiction. Along each curve in
Fig.~\ref{fig:H2-eta}, what is fixed is not $\Phi_-^2$ but
$\Phi_+^2$. If $h^2$ is increased from zero with $\Phi_-^2$ fixed then
Fig.~\ref{fig:H2+eta} says that $\Phi_+^2=\Phi_-^2/\eta_+^2$ should
increase. In Fig.~\ref{fig:H2-eta} this means that we have to move from
a curve with the initial $\Phi_+^2$ to different curves with larger
$\Phi_+^2$. Hence, if the value of $\eta_-$ were the same then this
would inevitably increase the value of $h^2$ from zero. Actually, since
$\eta_-$ is the reciprocal of $\eta_+$, Fig.~\ref{fig:H2+eta} says that
$\eta_-$ should also increase. Therefore, when $h^2$ is increased from
zero, there are two competitive effects which Fig.~\ref{fig:H2+eta}
implies: (i) the increase of $\Phi_+^2$; (ii) the increase of
$\eta_-$. The effect (i) tends to increase $h^2$ from zero but the
effect (ii) tends to decrease $h^2$. The result of the competition is
that the effect (i) wins. Hence, the negativity of the slopes of the
curves both in Fig.~\ref{fig:H2-eta} and Fig.~\ref{fig:H2+eta} near the
horizontal axis is consistent with each other.

The intersection of each curve in Fig.~\ref{fig:H2+eta} with the
horizontal axis defines the value $\eta_+^{(0)}(\Phi_-^2)$ of $\eta_+$
corresponding to $h_+^2=0$. Thus, 
%
\begin{equation}
 \sigma_+^{(0)}(\Phi_-^2) \equiv 
  \left[2\pi-(2\pi-\delta_-)\eta_+^{(0)}(\Phi_-^2)\right]\cdot M^4 
\end{equation}
is the critical value of tension $\sigma_+$ for which the effective
cosmological constant on the brane vanishes. If
$\sigma_+>\sigma_+^{(0)}(\Phi_-^2)$ (or
$\sigma_+<\sigma_+^{(0)}(\Phi_-^2)$) then the effective cosmological
constant on the brane is positive (or negative, respectively).

\begin{figure}
 \centering\leavevmode\epsfysize=8cm \epsfbox{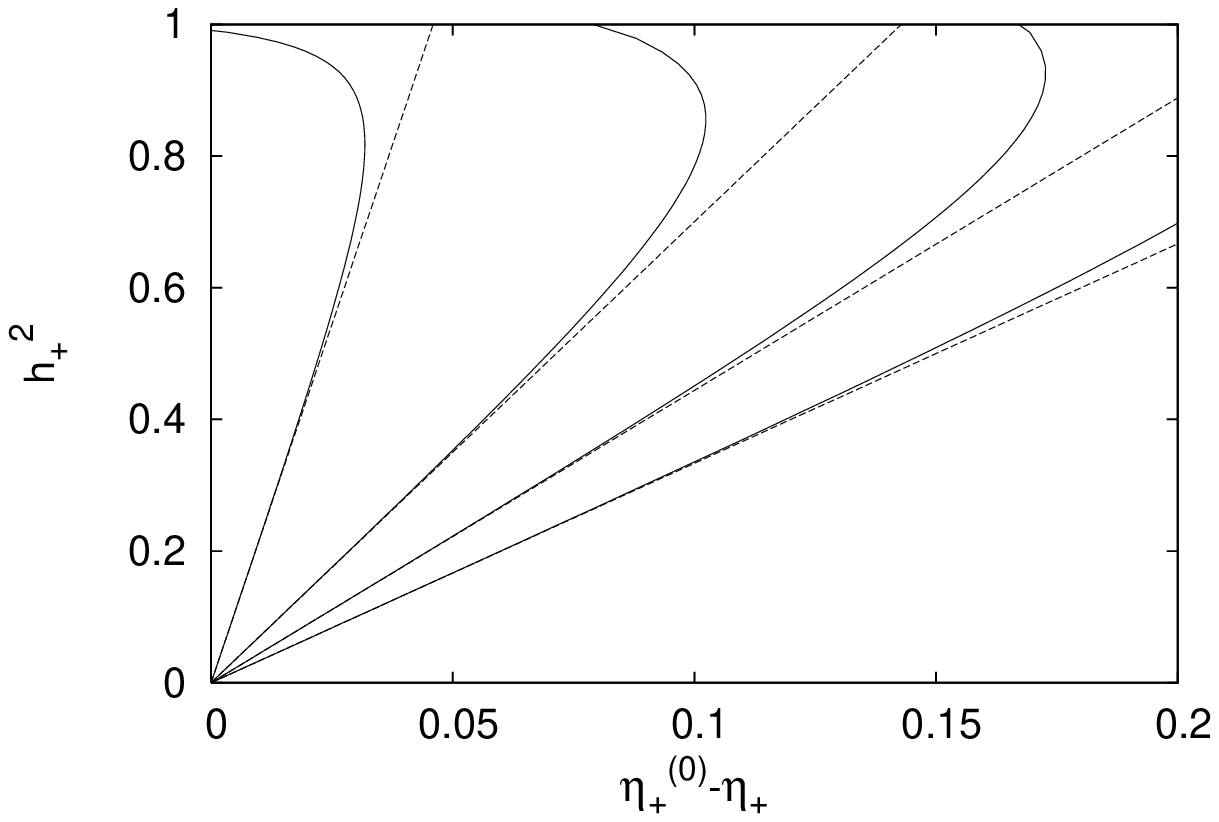}
 \caption{\label{fig:H+squared}
 Solid lines are curves shown in Fig.~\ref{fig:H2+eta} in a different
 coordinate. The vertical axis is again $h_+^2$ but the horizontal axis
 is now $\eta_+^{(0)}(\Phi_-^2)-\eta_+$, where $\eta_+^{(0)}(\Phi_-^2)$
 is the value of $\eta_+$ corresponding to $h_+^2=0$ for a given value
 of $\Phi_-^2$. The value of $\Phi_-^2$ for each curve is $0.02$,
 $0.08$, $0.14$, $0.2$ from left to right. Dot lines represent straight
 lines passing through the origin with the slope $g_+^{(0)}(\Phi_-^2)$
 for each value of $\Phi_-^2$. These graphs confirm that $G_{N+}$
 defined by (\ref{eqn:NewtonConstBrane}) indeed has the meaning of 
 $4$-dimensional Newton's constant on the brane at $r=r_+$.
 }
\end{figure}
The positivity of low energy effective Newton's constant is
highlighted in Fig.~\ref{fig:H+squared}, where the vertical axis is
again $h_+^2$ but the horizontal axis is now
$\eta_+^{(0)}(\Phi_-^2)-\eta_+$. The positive slope of the curve near
the origin clearly defines low energy effective Newton's constant
since it determines how much the Hubble expansion rate changes as the
brane tension changes. The dot lines in Fig.~\ref{fig:H+squared} are
straight lines passing through the origin with the slope
$g_+^{(0)}(\Phi_-^2)$, where $g_+^{(0)}(\Phi_-^2)$ is the value of $g_+$
with $h_+^2=0$ for a given value of $\Phi_-^2$. It is easily seen that
the solid lines and the dot lines in Fig.~\ref{fig:H+squared} come in
contact with each other at the origin. This fact implies that 
%
\begin{equation}
 h_+^2 = g_+^{(0)}\cdot(\eta_+^{(0)}-\eta_+)
  + O((\eta_+^{(0)}-\eta_+)^2).
  \label{eqn:FriedmannEq+pre}
\end{equation}
By definition of $g_+$ and the formula (\ref{eqn:deficit-angle}), this
is equivalent to 
%
\begin{equation}
 H_+^2 = \frac{8\pi G_{N+}}{3}(\sigma_+-\sigma_+^{(0)})
  + O((\sigma_+-\sigma_+^{(0)})^2),
  \label{eqn:FriedmannEq+}
\end{equation}
which confirms that $G_{N+}$ defined in (\ref{eqn:NewtonConstBrane})
indeed has the meaning of $4$-dimensional Newton's constant on the
brane at $r=r_+$. The straight dot lines in Fig.~\ref{fig:H+squared}
are good approximation to the solid lines in the regime where $h_+^2$ is
sufficiently smaller than $1$. This suggests that the Friedmann equation
should be recovered at low energy where the Hubble expansion rate
squared $H_+^2$ on the brane is sufficiently smaller than the bulk
cosmological constant $\Lambda_{6}$ ($=10L^{-2}$). In the next 
subsection~\ref{subsec:lowEexpansion}, we shall show the relation
(\ref{eqn:FriedmannEq+}) analytically.

\subsection{Low energy expansion}
\label{subsec:lowEexpansion}

In this subsection we show the low-energy relations
(\ref{eqn:FriedmannEq-}) and (\ref{eqn:FriedmannEq+}) analytically,
where Newton's constant $G_{N\pm}$ is given by
(\ref{eqn:NewtonConstBrane}). For this purpose we shall perform 
perturbative expansion of relevant quantities and equations
w.r.t. $h^2$ ($=h_+h_-$) and show that (\ref{eqn:FriedmannEq-pre}) and
(\ref{eqn:FriedmannEq+pre}) indeed hold in the lowest order in the
perturbative expansion.

Let us first consider the case considered in
subsection~\ref{subsec:IRbrane}, where the phase transition takes place 
on the brane at $r=r_-$. In this case we expand $\alpha$ and $\eta_-$ as
%
\begin{equation}
 \alpha = \sum_{n=0}^{\infty}\alpha_-^{(n)}(\Phi_+^2)h^{2n}, \quad
  \eta_- = \sum_{n=0}^{\infty}\eta_-^{(n)}(\Phi_+^2)h^{2n}. 
\end{equation}
Since $\Phi_+$ is fixed during the phase transition, we formally
consider $\Phi_+$ as a quantity of order $O(h^0)$. It is straightforward
to expand (\ref{eqn:flux-squared}) and (\ref{eqn:ratio-angle})
w.r.t. $h^2$. In the order $O(h^0)$, we obtain
%
\begin{eqnarray}
 \Phi_+^2 & = & \frac{16\beta_2^{(0)}\beta_1^{(0)3}}
  {3\gamma_+^{(0)2}\alpha_-^{(0)4}},
  \nonumber\\
 \eta_-^{(0)} & = & \frac{\gamma_-^{(0)}}{\gamma_+^{(0)}\alpha_-^{(0)4}},
\end{eqnarray}
where $\beta_n^{(0)}$ and $\gamma_{\pm}^{(0)}$ are $\beta_n$ and
$\gamma_{\pm}$, respectively, with $\alpha$ replaced by
$\alpha_-^{(0)}$. On the other hand, the zeroth order part of
eq.~(\ref{eqn:WarpedVolume}) gives the value $g_-^{(0)}(\Phi_+^2)$ of
$g_-$ with $h_-^2=0$ for a given value of $\Phi_+^2$ as
%
\begin{equation}
 g_-^{(0)} = \frac{\gamma_+^{(0)}\alpha_-^{(0)3}}{2\beta_1^{(0)2}}.
\end{equation}

In the order $O(h^2)$, eq.~(\ref{eqn:flux-squared}) divided by
(\ref{eqn:ratio-angle}) gives an expression of $\alpha_-^{(1)}$ in terms
of $\alpha_-^{(0)}$ as 
%
\begin{equation}
 \frac{\alpha_-^{(1)}}{\alpha_-^{(0)}} = 
  \frac{\beta_1^{(0)2}A_-^{(0)}}{B_-^{(0)}},
\end{equation}
where
%
\begin{eqnarray}
 A_-^{(0)} & = & 3\alpha_-^{(0)3}+6\alpha_-^{(0)2}
  +9\alpha_-^{(0)}+6+3\alpha_-^{(0)-1}+2\alpha_-^{(0)-2}+\alpha_-^{(0)-3}, 
  \nonumber\\
 B_-^{(0)} & = & 15\alpha_-^{(0)6}+60\alpha_-^{(0)5}+150\alpha_-^{(0)4}
  +258\alpha_-^{(0)3}+357\alpha_-^{(0)2}+430\alpha_-^{(0)}+460
  \nonumber\\
 & & 
  +430\alpha_-^{(0)-1}+357\alpha_-^{(0)-2}+258\alpha_-^{(0)-3}
  +150\alpha_-^{(0)-4}+60\alpha_-^{(0)-5}+15\alpha_-^{(0)-6}. \nonumber\\
\end{eqnarray}
Here, as already explained, $\Phi_+$ was considered as a quantity of
order $O(h^0)$ in the derivation of the above expression of
$\alpha_-^{(0)}$. On the other hand, the reciprocal of
eq.~(\ref{eqn:ratio-angle}) in the order $O(h^2)$ gives 
%
\begin{equation}
 \eta_-^{(1)} = -\frac{2\beta_1^{(0)2}}{\gamma_+^{(0)}\alpha_-^{(0)4}}
  = -\frac{1}{g_-^{(0)}\alpha_-^{(0)}}. 
\end{equation}

Now, the definition of $\eta_-^{(0,1)}$ 
%
\begin{equation}
 \eta_- = \eta_-^{(0)} + \eta_-^{(1)}h^2 + O(h^4)
\end{equation}
is solved w.r.t. $h^2$ as
%
\begin{equation}
 h^2 = 
  \frac{\eta_--\eta_-^{(0)}}{\eta_-^{(1)}} + O(h^4),
\end{equation}
or equivalently
%
\begin{equation}
 h_-^2 = \frac{h^2}{\alpha}
  = g_-^{(0)}\cdot
  \left(\eta_-^{(0)}-\eta_-\right) + O(h_-^4).
\end{equation}
This proves (\ref{eqn:FriedmannEq-pre}). By the definition of $g_-$ and
the formula (\ref{eqn:deficit-angle}), this is equivalent to
(\ref{eqn:FriedmannEq-}).

It is easy to perform a similar analysis for the other case considered
in subsection~\ref{subsec:UVbrane}, where the phase transition takes
place on the brane at $r=r_+$, and to derive (\ref{eqn:FriedmannEq-pre})
and (\ref{eqn:FriedmannEq+}) analytically. In this case we expand
$\alpha$ and $\eta_+$ as 
%
\begin{equation}
 \alpha = \sum_{n=0}^{\infty}\alpha_+^{(n)}(\Phi_-^2)h^{2n}, \quad
  \eta_+ = \sum_{n=0}^{\infty}\eta_+^{(n)}(\Phi_-^2)h^{2n}. 
\end{equation}
Since $\Phi_-$ is fixed during the phase transition, we formally
consider $\Phi_-$ as a quantity of order $O(h^0)$. It is straightforward
to expand (\ref{eqn:flux-squared}) and (\ref{eqn:ratio-angle})
w.r.t. $h^2$. In the order $O(h^0)$, we obtain
%
\begin{eqnarray}
 \Phi_-^2 & = & \frac{16\beta_2^{(0)}\beta_1^{(0)3}\alpha_+^{(0)4}}
  {3\gamma_-^{(0)2}},
  \nonumber\\
 \eta_+^{(0)} & = & \frac{\gamma_+^{(0)}\alpha_+^{(0)4}}{\gamma_-^{(0)}},
\end{eqnarray}
where $\beta_n^{(0)}$ and $\gamma_{\pm}^{(0)}$ are $\beta_n$ and
$\gamma_{\pm}$, respectively, with $\alpha$ replaced by
$\alpha_+^{(0)}$. On the other hand, the zeroth order part of
eq.~(\ref{eqn:WarpedVolume}) gives the value $g_+^{(0)}(\Phi_+^2)$ of 
$g_+$ with $h_+^2=0$ for a given value of $\Phi_-^2$ as
%
\begin{equation}
 g_+^{(0)} = \frac{\gamma_-^{(0)}}{2\beta_1^{(0)2}\alpha_+^{(0)3}}.
\end{equation}

In the order $O(h^2)$, eq.~(\ref{eqn:flux-squared}) multiplied by
(\ref{eqn:ratio-angle}) gives an expression of $\alpha_+^{(1)}$ in terms
of $\alpha_+^{(0)}$ as 
%
\begin{equation}
 \frac{\alpha_+^{(1)}}{\alpha_+^{(0)}} = 
  \frac{\beta_1^{(0)2}A_+^{(0)}}{B_+^{(0)}},
\end{equation}
where
%
\begin{eqnarray}
 A_+^{(0)} & = & 3\alpha_+^{(0)-3}+6\alpha_+^{(0)-2}
  +9\alpha_+^{(0)-1}+6+3\alpha_+^{(0)}+2\alpha_+^{(0)2}+\alpha_+^{(0)3},
  \nonumber\\
 B_+^{(0)} & = & 15\alpha_+^{(0)-6}+60\alpha_+^{(0)-5}+150\alpha_+^{(0)-4}
  +258\alpha_+^{(0)-3}+357\alpha_+^{(0)-2}+430\alpha_+^{(0)-1}+460
  \nonumber\\
 & & 
  +430\alpha_+^{(0)}+357\alpha_+^{(0)2}+258\alpha_+^{(0)3}
  +150\alpha_+^{(0)4}+60\alpha_+^{(0)5}+15\alpha_+^{(0)6}. 
\end{eqnarray}
On the other hand, eq.~(\ref{eqn:ratio-angle}) in the order $O(h^2)$
gives 
%
\begin{equation}
 \eta_+^{(1)} = -\frac{2\beta_1^{(0)2}\alpha_+^{(0)4}}{\gamma_-^{(0)}}
  = -\frac{\alpha_+^{(0)}}{g_+^{(0)}}. 
\end{equation}

Now, the definition of $\eta_+^{(0,1)}$ 
%
\begin{equation}
 \eta_+ = \eta_+^{(0)} + \eta_+^{(1)}h^2 + O(h^4)
\end{equation}
is solved w.r.t. $h^2$ as
%
\begin{equation}
 h^2 = 
  \frac{\eta_+-\eta_+^{(0)}}{\eta_+^{(1)}} + O(h^4),
\end{equation}
or equivalently
%
\begin{equation}
 h_+^2 = h^2\alpha
  = g_+^{(0)}\cdot
  \left(\eta_+^{(0)}-\eta_+\right) + O(h_+^4).
\end{equation}
This proves (\ref{eqn:FriedmannEq+pre}). By the definition of $g_+$ and
the formula (\ref{eqn:deficit-angle}), this is equivalent to
(\ref{eqn:FriedmannEq+}).

It is straightforward to extend the analysis in this subsection to to
higher oder in the expansion w.r.t. $h^2$ and to interpret the result as
higher-order corrections to the effective Friedmann equation. In
Appendix~\ref{app:rho2-correction} the results in the order $O(h^4)$ is 
summarized and it is shown that higher order corrections can be ignored
when the Hubble expansion rate $H_{\pm}$ on a brane is sufficiently
lower than the bulk curvature scale $L^{-1}$.

\section{Summary and discussion}
\label{sec:summary}

We have considered a $6$-dimensional model of warped flux
compactification and analyzed gravity sourced by a $3$-brane in a
simplified setup. This setup includes a warped geometry,
compactification, a flux, and one or two $3$-brane(s). We have
considered a situation where the tension of a $3$-brane changes by a
phase transition on the brane. Assuming that the tension is almost
constant in deep inside the old and new phases, we have investigated the
relation between the change of tension and the change of the Hubble
expansion rate on the brane. The relation is the same as that inferred
by the $4$-dimensional Einstein equation, provided that Newton's
constant is given by the formula (\ref{eqn:NewtonConstBrane}).

We have explicitly seen how the induced geometry changes according to a 
change of brane source. With the static ansatz for the bulk, the bulk
geometry is uniquely determined by the brane tensions and the value of
the conserved flux. As a consequence, the Hubble expansion rate induced
on the brane is determined. Provided that the solutions considered in
this paper are dynamically stable~\cite{Stability}, this result implies
that, when a brane tension changes, the bulk geometry should evolve
toward the corresponding unique configuration. This process determines
the evolution of the bulk geometry and, thus, the induced geometry on
the brane. In this way, the induced geometry responses to the brane
source.

Although the situation we have considered is rather restrictive, the
physical picture we have obtained seems more general. Provided that all
moduli are stabilized, when a brane source changes slowly compared to
the time scale of the moduli dynamics, the bulk geometry should quickly
settle to a configuration which is determined by the boundary condition,
i.e. the brane source(s), values of conserved quantities and the
regularity of the other region of the extra dimensions. As a consequence
of the change of the bulk geometry, the induced geometry on the brane
responses to the evolution of the matter source on the brane.

Suppose that all moduli are stabilized in a ($4+n$)-dimensional warped
compactification. From the viewpoint of the effective field theory with
massive moduli integrated out, it is in general expected that the
Einstein gravity should be recovered at low energy~\footnote{
The stress-energy tensor in the $4$-dimensional effective Einstein
equation must be calculated from a $4$-dimensional effective action
after integrating out all massive moduli. The effective stress-energy
tensor defined in this way is not necessarily the same as the surface
stress-energy tensor (the so called Lanczos tensor~\cite{Lanczos}
in Israel's junction condition~\cite{Israel} for codimension $1$
cases). If moduli are not stabilized, it is a non-trivial issue whether
a $4$-dimensional theory is recovered. See
eg.~\cite{Shiromizu,Koyama-Koyama} for this issue in the context of 
Randall-Sundrum type braneworlds. 
} 
and that $4$-dimensional Newton's constant should be given by the
formula (\ref{eqn:GN-formula}):  
%
\begin{equation}
 \frac{1}{8\pi G_N} = 
  (M_{4+n})^{2+n}\int d^ny 
  \sqrt{\gamma}\left[\frac{r(y)}{r(y_0)}\right]^2. \nonumber\\
\end{equation}
(See the first paragraph of Sec.~\ref{sec:Friedmann-eq} for the
reasoning leading to this expectation.) Here, $M_{4+n}$ is the
($4+n$)-dimensional Planck mass, $r^2(y)$ is the warp factor of mass
dimension $-2$ depending on the coordinates of compact extra dimensions
$y^i$, $\gamma_{ij}dy^idy^j$ is the metric of the extra dimensions (see 
eq.~(\ref{eqn:warped-metric-general})) and the brane source is supposed
to be localized at $y^i=y^i_0$. Although the effective field theory
approach is usually a powerful and useful tool to obtain a correct
answer quickly, the result must be checked by independent methods. We
expect that this formula should hold at low energy in a wide class of
warped compactification if all moduli are stabilized and made massive.

The formula (\ref{eqn:GN-formula}) was already proved to be correct in a
Randall-Sundrum type, codimension $1$ braneworlds with radion
stabilization~\cite{Tanaka-Montes,CGK,Mukohyama-Kofman,Mukohyama-HDterm,Kudoh-Tanaka,CGR,KKOP}~\footnote{See
the footnote after (\ref{eqn:GN-formula}).}. The result of the present
paper confirms the same formula for codimension $2$ braneworlds in a
simplified situation.

Future subjects include the analysis of dynamical
stability~\cite{Stability}, the recovery of the $4$-dimensional
linearized Einstein equation for inhomogeneous perturbations, evolution
of FRW universe on the brane, extensions to systems with a dilaton, and
so on.

The simple $6$-dimensional setup in this paper at the very least
provides non-trivial evidence for validity of the expected formula
(\ref{eqn:GN-formula}) for effective Newton's constant in warped flux
compactification. Thus, let us now briefly discuss application of the
formula (\ref{eqn:GN-formula}) to cosmology with type IIB warped flux
compactification considered in
refs.~\cite{KKLT,KKLMMT,BBCEGKLQ,InflationKKLTsetup,Polchinski,Mukohyama2005}.
The $10$-dimensional geometry is of the form 
%
\begin{equation}
 ds_{10}^2 = h^{-1/2}(\tau) g^{(4)}_{\mu\nu}dx^{\mu}dx^{\nu}
  + h^{1/2}(\tau)\gamma_{mn}d\psi^md\psi^n,
\end{equation}
where $\gamma_{mn}$ is the metric of a $6$-dimensional compact geometry
(Calabi-Yau space) depending on the coordinates $\psi^m$ 
($m=5,\cdots, 10$) of extra dimensions, $h(\tau)$  is a function of the
radial coordinate $\tau$ ($\in\{\psi^m\}$) of the compact geometry, and
$g^{(4)}_{\mu\nu}$ is a $4$-dimensional metric. It is supposed that the
function $h(\tau)$ is of order $1$ in the bulk of the compact geometry
but that there is a region called a warped throat where $h(\tau)$
becomes exponentially large. In the warped throat region the warp
factor $h^{-1/2}(\tau)$ becomes exponentially small.

In most of 
refs.~\cite{KKLMMT,BBCEGKLQ,InflationKKLTsetup,Polchinski,Mukohyama2005}
it was (implicitly) assumed that what drives (or at least affects) the
$4$-dimensional cosmology is a brane in the throat region but that our
$4$-dimensional world is somewhere in the Calabi-Yau space where the 
warp factor $h^{-1/2}(\tau)$ is of order unity. In other words, the
physical metric of our $4$-dimensional world is $g^{(4)}_{\mu\nu}$ up to
a normalization constant of order unity, while the induced metric on the
brane is not $g^{(4)}_{\mu\nu}$ but 
%
\begin{equation}
 g^{(b)}_{\mu\nu} = h^{-1/2}(\tau_0)\cdot g^{(4)}_{\mu\nu},
  \label{eqn:g-brane}
\end{equation}
where $\tau_0$ is the value of the radial coordinate $\tau$ at the
position of the brane. At first sight, it might seem non-trivial how the
brane at $\tau=\tau_0$ drives the cosmology of our $4$-dimensional world
$g^{(4)}_{\mu\nu}$ since the warp factor $h^{-1/2}(\tau_0)$ is
exponentially small. The approach adopted in the literature is to write
down the brane action in terms of $g^{(4)}_{\mu\nu}$ and couple it to
the Einstein gravity in the frame $g^{(4)}_{\mu\nu}$, assuming that
Newton's constant $G^{(4)}_N$ is given by the Kaluza-Klein like formula
%
\begin{equation}
 \frac{1}{8\pi G^{(4)}_N} = M_{10}^8V_6,
  \label{eqn:g-KK}
\end{equation}
where $V_6$ is the volume of the Calabi-Yau space.

We can justify this approach by using the formula
(\ref{eqn:GN-formula}). First, the induced geometry on the brane is, 
as already stated, the conformally transformed metric
$g^{(b)}_{\mu\nu}$ given in (\ref{eqn:g-brane}). Thus, the stress
energy tensor $T^{(b)}_{\mu\nu}$ in the brane frame
$g^{(b)}_{\mu\nu}$ and the stress energy tensor $T^{(4)}_{\mu\nu}$
in the frame $g^{(4)}_{\mu\nu}$ are related to each other as 
%
\begin{equation}
 T^{(b)\nu}_{\mu} = h(\tau_0)\cdot T^{(4)\nu}_{\mu},
\end{equation}
where $\tau_0$ is again the value of the radial coordinate $\tau$ at the
position of the brane and it has been assumed that the value of
$h(\tau)$ at the position of our $4$-dimensional world is normalized to
$1$. Now, the formula (\ref{eqn:GN-formula}) applied to this setup says
that we can use the effective Einstein equation of the form 
%
\begin{equation}
 G^{(b)\nu}_{\mu} = 8\pi G^{(b)}_NT^{(b)\nu}_{\mu},
  \label{eqn:eq-on-brane}
\end{equation}
where $G^{(b)\nu}_{\mu}$ is the Einstein tensor of the brane metric
$g^{(b)}_{\mu\nu}$ and Newton's constant $G^{(b)}_N$ on the brane is
given by 
%
\begin{equation}
 \frac{1}{8\pi G^{(b)}_N} = M_{10}^8\int d^6\psi
  h^{3/2}(\tau)\sqrt{\gamma}\cdot
  \left(\frac{h(\tau)}{h(\tau_0)}\right)^{-1/2}.
\end{equation}
By the assumption that the function $h(\tau)$ is of order unity in the
bulk of extra dimensions, this is reduced to 
%
\begin{equation}
 \frac{1}{8\pi G^{(b)}_N} \simeq h^{1/2}(\tau_0)M_{10}^8V_6
  \simeq \frac{h^{1/2}(\tau_0)}{8\pi G^{(4)}_N},
\end{equation}
where $V_6$ is again the volume of the Calabi-Yau space. This result
implies that the effective Einstein equation on the brane
(\ref{eqn:eq-on-brane}) is equivalent to 
%
\begin{equation}
 G^{(4)\nu}_{\mu} = 8\pi G^{(4)}_NT^{(4)\nu}_{\mu},
\end{equation}
where we have used the conformal transformation
%
\begin{equation}
 G^{(b)\nu}_{\mu} =  h^{1/2}(\tau_0)G^{(4)\nu}_{\mu}
\end{equation}
between Einstein tensors $G^{(b)\nu}_{\mu}$ and $G^{(4)\nu}_{\mu}$ of
$g^{(b)}_{\mu\nu}$ and $g^{(4)}_{\mu\nu}$, respectively. Therefore, we
have justified the approach adopted in the literature by using the
formula (\ref{eqn:GN-formula}). The $4$-dimensional fields resulting
from the brane action indeed couple to the $4$-dimensional metric
$g^{(4)}_{\mu\nu}$ with Newton's constant $G^{(4)}_N$ given by the
Kaluza-Klein like formula (\ref{eqn:g-KK}).

\section*{Acknowledgements}

We would like to thank Andrei Frolov, Yoshiaki Himemoto, Lev Kofman,
Kazuya Koyama, Hideaki Kudoh, Tetsuya Shiromizu, Jiro Soda, Keitaro
Takahashi and Takahiro Tanaka for useful discussions and/or comments. We
are grateful to Katsuhiko Sato for continuing encouragement. Y.~S. and
H.~Y. are supported by JSPS.

\appendix{Appendices}

\subsection{The $\alpha\to 1$ limit}
\label{app:alphato1}

In the limit $\alpha\equiv r_-/r_+\to 1$, the coordinate distance
between two branes vanishes and, thus, the bulk geometry appears to
collapse. However, as we shall see below, the proper distance between
$r=r_-$ and $r_+$ does not vanish and the geometry of extra dimensions
remains regular.

When $\alpha$ is sufficiently close to $1$, the function $f(r)$ between
$r_{\pm}$ is approximated by 
%
\begin{equation}
 f(r) \simeq a^2 - b^2(r-r_0)^2,
\end{equation}
where $r_0$, $a$ and $b$ are positive constants and the $\alpha\to 1$ 
limit corresponds to sending $a\to +0$ with $b$ finite. The bulk
coordinate $r$ is restricted to the interval $-a/b\leq r-r_0\leq a/b$. 
With the new coordinate system ($\theta$, $\varphi$) defined by 
%
\begin{equation}
 \theta = \cos^{-1}\left[\frac{b}{a}(r-r_0)\right], 
  \quad \varphi = ab\phi, \label{eqn:coord-alpha-1-limit}
\end{equation}
the metric of the extra dimension becomes that of a round sphere with 
radius $1/b$: 
%
\begin{equation}
 \frac{dr^2}{f(r)}+f(r)d\phi^2 
  \simeq \frac{1}{b^2}(d\theta^2+\sin^2\theta d\varphi^2). 
\end{equation}

Evidently, the coordinate $\theta$ runs over the full interval
$[0,\pi]$. On the other hand, the period of the coordinate $\varphi$ 
appears to collapse since the coefficient $ab$ in the definition
(\ref{eqn:coord-alpha-1-limit}) of $\varphi$ vanishes in the 
$\alpha\to 1$ limit. Actually, this is not the case. The ``surface
gravity'' $\kappa_{\pm}$ defined in (\ref{eqn:def-rho-kappa}) is
$\kappa_+\simeq \kappa_- \simeq ab$ and the period of the old coordinate
is, thus, $\Delta\phi\simeq (2\pi-\delta_{\pm})/(ab)$. Therefore, the
new coordinate $\varphi$ has the period 
%
\begin{equation}
 \Delta\varphi = ab\Delta\phi \simeq 2\pi -\delta_+ 
  \simeq 2\pi -\delta_-,
\end{equation}
which is indeed finite. Thus, the geometry of extra dimensions is
nothing but a round sphere with a deficit angle
$\delta_+\simeq\delta_-$, i.e. a football-shaped extra-dimensions
considered in \cite{Carroll-Guica,Garriga-Porrati,Navarro}. In those
papers a $Z_2$ symmetry is assumed so that tensions of two $3$-branes
are identical. With the $Z_2$ symmetry, $\eta_{\pm}$ is always $1$ but 
$\Phi_{\pm}=\Phi/(2\pi-\delta_{\pm})$ changes when the brane tension
changes. Accordingly, as can be seen from Figs.~\ref{fig:H2-eta} and
\ref{fig:H2+eta}, the Hubble expansion rate on the brane changes when
the brane tension changes.

The warp factor $r$ becomes constant in the $\alpha\to 1$ limit and,
thus, the $6$-dimensional solution in the $\alpha\to 1$ limit
corresponds to an unwarped flux compactification of the type considered
by Arkani-Hamed, Dimopoulos and Dvali~\cite{ADD}.

\subsection{The $h\to 0$ limit}
\label{app:Hto0}

The limit $h\to 0$ appears to be singular, but it is not. Indeed, by the
coordinate change
%
\begin{equation}
 r \to \frac{\bar{r}}{h}, \quad \phi \to h\bar{\phi}, \quad 
  ds_4^2 \to h^2d\bar{s}_4^2, \quad
  \mu \to \frac{\bar{\mu}}{h^5}, \quad b \to \frac{\bar{b}}{h^4}, 
\end{equation}
the $h\to 0 $ limit of the bulk solution becomes a locally regular
form: 
%
\begin{eqnarray}
 ds_6^2 & = & \bar{r}^2d\bar{s}_4^2 
  + \frac{d\bar{r}^2}{\bar{f}(\bar{r})} + \bar{f}(\bar{r})d\bar{\phi}^2, 
  \nonumber\\
 A_Mdx^M & = & \frac{\bar{b}}{3\bar{r}^3}d\bar{\phi},
\end{eqnarray}
where 
%
\begin{equation}
 \bar{f}(\bar{r}) = -\frac{\Lambda_6}{10}\bar{r}^2
  -\frac{\bar{\mu}}{\bar{r}^3}
  -\frac{\bar{b}^2}{12\bar{r}^6},
\end{equation}
and $d\bar{s}_4^2$ is the metric of the $4$-dimensional Minkowski
spacetime. This configuration of course satisfies the $6$-dimensional
Einstein equation and is related by a double Wick rotation to a
topological black hole with the horizon topology ${\bf R}^4$.

The new coordinate $\bar{\phi}$ is identified as
%
\begin{equation}
 \bar{\kappa}_{\pm}\bar{\phi} \sim 
  \bar{\kappa}_{\pm}\bar{\phi} + (2\pi -\delta_{\pm}),
\end{equation}
where
%
\begin{equation}
   \bar{\kappa}_{\pm} \equiv \mp \frac{1}{2}\bar{f}'(\bar{r}_{\pm}). 
\end{equation}
This means that the period of $\bar{\phi}$ is finite if the brane
tensions are finite. Therefore, the geometry remains regular in the
$h\to 0$ limit.

\subsection{Higher-order correction}
\label{app:rho2-correction}

It is straightforward to extend the analysis in
subsection~\ref{subsec:lowEexpansion} to higher order in the expansion
w.r.t. $h^2$ and to interpret the result as higher-order corrections to 
the effective Friedmann equation. In this appendix we just summarize the
result up to the order $O(h^4)$. 

The result up to the order $O(h^4)$ is summarized as
%
\begin{equation}
 \left[ 1 - \frac{h_{\pm}^2}{h_{*\pm}^2} + O(h_{\pm}^4)\right]
  h_{\pm}^2  = g_{\pm}^{(0)}\cdot
  \left(\eta_{\pm}^{(0)}-\eta_{\pm}\right),
\end{equation}
where
%
\begin{equation}
 h_{*\pm}^2 = \frac{B_{\pm}^{(0)}}{\beta_1^2A_{\pm}^{(0)}}\alpha_{\pm}^{(0)\pm 1}. 
\end{equation}
This is equivalent to 
%
\begin{equation}
 \left[ 1 - \frac{H_{\pm}^2}{H_{*\pm}^2}
   + O\left((H^2_{\pm}/\Lambda_6)^2\right) \right] H_{\pm}^2
  = \frac{8\pi G_{N\pm}}{3}(\sigma_{\pm}-\sigma_{\pm}^{(0)}),
\end{equation}
where $H_{*\pm}^2=h_{*\pm}^2/L^2$, and includes a higher order
correction to the effective Friedmann equation suppressed by
$H_{*\pm}$. It is easy to show that
%
\begin{equation}
 5 < L^2H_{*+}^2 \leq \left(\frac{10}{3}\right)^2,
  \quad
 \left(\frac{10}{3}\right)^2 \leq L^2H_{*-}^2 < \infty.
\end{equation}
This means that higher order corrections can be ignored when the Hubble
expansion rate $H_{\pm}$ on a brane is sufficiently lower than the bulk
curvature scale $L^{-1}$.



\begin{thebibliography}{99}
 \bibitem{KKLT}
         S.~Kachru, R.~Kallosh, A.~Linde and S.~P.~Trivedi,
         Phys. Rev. {\bf D68}, 046005 (2003) [hep-th/0301240].
 \bibitem{EGQ}
	 C.~Escoda, M.~Gomez-Reino and F.~Quevedo, JHEP {\bf 0311}, 065
	 (2003) [hep-th/0307160]. 
 \bibitem{BKQ}
	 C.~P.~Burgess, R.~Kallosh and F.~Quevedo JHEP {\bf 0310}, 056
	 (2003) [hep-th/0309187].
 \bibitem{Silverstein}
         E.~Silverstein, ``(A)dS Backgrounds from Asymmetric
         Orientifolds'', hep-th/0106209; 
         A.~Maloney, E.~Silverstein and A.~Strominger, ``de Sitter Space
         in Noncritical String Theory'', hep-th/0205316. 
 \bibitem{BCK}
	 M.~Becker, G.~Curio and A.~Krause, Nucl. Phys. {\bf B693}, 223
	 (2004) [hep-th/0403027]. 
 \bibitem{Townsend-Wohlfarth}
         P.~K.~Townsend and M.~N.~R.~Wohlfarth, Phys. Rev. Lett. 
         {\bf 91}, 061302 (2003) [hep-th/0303097].
 \bibitem{Ohta}
         N.~Ohta, Phys. Rev. Lett. {\bf 91}, 061303 (2003)
         [hep-th/0303238];  
         Prog. Theor. Phys. {\bf 110}, 269 (2003) [hep-th/0304172];
         Int.~J.~Mod.~Phys. {\bf A20}, 1 (2005) [hep-th/0411230]. 
 \bibitem{Roy}
	 S.~Roy, ``Accelerating cosmologies from M/String theory
	 compactifications'', Phys.~Lett. {\bf B567}, 322 (2003)
	 [hep-th/0304084].
 \bibitem{Neupane}
	I.~P.~Neupane and D.~L.~Wiltshire, ``Accelerating cosmologies
	 from compactification with a twist'', hep-th/0502003; 
	 ``Cosmic acceleration from M-theory on twisted spaces'',
	 hep-th/0504135. 
 \bibitem{BBK}
	 K.~Becker, M.~Becker and A.~Krause,
	 Nucl. Phys. {\bf B715}, 349 (2005) [hep-th/0501130].
 \bibitem{KS}
         I.~R.~Klebanov and M.~J.~Strassler, JHEP {\bf 0008}, 052
         (2000) [hep-th/0007191].
 \bibitem{GKP}
         S.~B.~Giddings, S.~Kachru, J.~Polchinski, Phys. Rev. {\bf D66},
         106006 (2002) [hep-th/0105097].
 \bibitem{Witten}
	 E.~Witten, Nucl. Phys. {\bf B474}, 343 (1996)
	 [hep-th/9604030]. 
 \bibitem{KKLMMT}
         S.~Kachru, R.~Kallosh, A.~Linde, J.~Maldacena, L.~McAllister,
         S.~P.~Trivedi, JCAP {\bf 0310}, 013 (2003) [hep-th/0308055].
 \bibitem{BBCEGKLQ}
	 J.~J.~Blanco-Pillado, C.~P.~Burgess, J.~M.~Cline, C.~ Escoda,
	 M.~Gomez-Reino, R.~Kallosh, A.~Linde and F.~Quevedo, 
	 JHEP {\bf 0411}, 063 (2004) [hep-th/0406230]. 
\bibitem{InflationKKLTsetup}
	 A.~Buchel and Radu Roiban, Phys. Lett. {\bf B590}, 284 (2004)
	 [hep-th/0311154]; 
	 M.~Berg and M.~Haack, Phys. Rev. {\bf D71}, 026005 (2005)
	 [hep-th/0404087]; 
	 C.~P.~Burgess, J.~M.~Cline, H.~Stoica and F.~Quevedo, JHEP 
	 {\bf 0409}, 033 (2004) [hep-th/0403119];
	 O.~DeWolfe, S.~Kachru and H.~Verlinde, JHEP {\bf 0405}, 017
	 (2004) [hep-th/0403123];
	  N.~Iizuka and S.~P.~Trivedi, Phys. Rev. {\bf D70}, 043519
	 (2004) [hep-th/0403203];
	 A.~Buchel and A.~Ghodsi, Phys. Rev. {\bf D70}, 126008 (2004)
	 [hep-th/0404151]; 
	 H.~Firouzjahi and S.~H.~Tye, JCAP {\bf 0503}, 009 (2005)
	 [hep-th/0501099];
	 J.~P.~Hsu, R.~Kallosh and S.~Prokushkin, JCAP {\bf 0312}, 009
	 (2003) [hep-th/0311077];
	 H.~Firouzjahi and S.~H.~Tye, Phys. Lett. {\bf B584}, 147 (2004)
	 [hep-th/0312020];
	 J.~P.~Hsu and R.~Kallosh, JHEP {\bf 0404}, 042 (2004)
	 [hep-th/0402047]; 
	 F.~Koyama, Y.~Tachikawa and T.~Watari, Phys. Rev. {\bf D69},
	 106001 (2004), Erratum-ibid. {\bf D70}, 129907 (2004)
	 [hep-th/0311191];
	 L.~McAllister,  ``An Inflaton Mass Problem in String Inflation
	 from Threshold Corrections to Volume Stabilization'',
	 hep-th/0502001; 
	 K.~Dasgupta, J.~P.~Hsu, R.~Kallosh, A.~Linde and M.~Zagermann,
	 JHEP {\bf 0408}, 030 (2004) [hep-th/0405247];
	 P.~Chen, K.~Dasgupta, K.~Narayan, M.~Shmakova and M.~Zagermann,
	 ``Brane inflation, solitons and cosmological solutions: I'', 
	 hep-th/0501185;
	 Yun-Song Piao, ``D/antiD Dark Energy in String Warped
	 Compactification'', gr-qc/0502006;
	 D.~Cremades, F.~Quevedo and A.~Sinha, ``Warped Tachyonic
	 Inflation in Type IIB Flux Compactifications and the
	 Open-string Completeness Conjecture'', hep-th/0505252. 
  \bibitem{Polchinski}
	 J.~Polchinski, ``Introduction to Cosmic F- and D-strings'',
	 hep-th/0412244 and references therein. 
 \bibitem{Mukohyama2005}
	 S.~Mukohyama, ``$\bar{D}$-brane as Dark Matter in Warped String
	 Compactification'', hep-th/0505042. 
 \bibitem{Dvali-Tye}
	 G.~R.~Dvali and S.~H.~Tye, Phys. Lett. {\bf B450}, 72 (1999)
	 [hep-ph/9812483]. 
 \bibitem{Quevedo}
	 F.~Quevedo, Class. Quant. Grav. {\bf 19}, 5721 (2002)
	 [hep-th/0210292] and references therein. 
 \bibitem{RS2}
	 L.~Randall and R.~Sundrum, Phys. Rev. Lett. {\bf 83}, 4690
	 (1999) [hep-th/9906064]. 
 \bibitem{Garriga-Tanaka}
	 J.~Garriga and T.~Tanaka, Phys. Rev. Lett. {\bf 84}, 2778
	 (2000) [hep-th/9911055]. 
 \bibitem{SSM}
	 M.~Sasaki, T.~Shiromizu and K.~Maeda, Phys. Rev. {\bf D62},
	 024008 (2000) [hep-th/9912233]. 
 \bibitem{GKR}
	 S.~B.~Giddings, E.~Katz, and L.~Randall, J. High Energy
	 Phys. {\bf 03}, 023 (2000) [hep-th/0002091].
 \bibitem{Karch-Randall}
	 A.~Karch and L.~Randall, JHEP {\bf 0105}, 008 (2001)
	 [hep-th/0011156]. 
 \bibitem{SMS}
	 T.~Shiromizu, K.~Maeda and M.~Sasaki, Phys. Rev. {\bf D62},
	 024012 (2000) [gr-qc/9910076]. 
 \bibitem{CGKT}
	 C.~Csaki, M.~Graesser, C.~F.~Kolda and J.~Terning, Phys. Lett.
	 {\bf B462}, 34 (1999) [hep-ph/9906513].
 \bibitem{CGS}
	 J.~M.~Cline, C.~Grojean and G.~Servant, Phys. Rev. Lett.
	 {\bf 83} 4245 (1999) [hep-ph/9906523].
 \bibitem{FTW}
	 E.~E.~Flanagan, S.~H.~H.~Tye, I.~Wasserman, Phys. Rev. {\bf D62},
	 044039 (2000) [hep-ph/9910498].
 \bibitem{BDEL}
	 P.~Bin\'{e}truy, C.~Deffayet, U.~Ellwanger and D.~Langlois,
	 Phys. Lett. {\bf B477}, 285 (2000) [hep-th/9910219].
 \bibitem{Mukohyama2000a}
	 S.~Mukohyama, Phys. Lett. {\bf B473}, 241 (2000) [hep-th/9911165].
 \bibitem{Kraus}
	 P.~Kraus, JHEP {\bf 9912}, 011 (1999) [hep-th/9910149].
 \bibitem{Ida}
	 D.~Ida, JHEP {\bf 0009}, 014 (2000) [gr-qc/9912002].
 \bibitem{MSM}
	 S.~Mukohyama, T.~Shiromizu and K.~Maeda, Phys. Rev. {\bf D62}, 024028
	 (2000), Erratum-ibid. {\bf D63}, 029901 (2001) [hep-th/9912287].
 \bibitem{RS1}
	 L.~Randall and R.~Sundrum, Phys. Rev. Lett. {\bf 83}, 3370
	 (1999) [hep-ph/9905221]. 
 \bibitem{GW}
	 W.~D.~Goldberger and M.~B.~Wise, Phys. Rev. Lett. {\bf 83},
	 4922 (1999) [hep-th/0007065]. 
 \bibitem{Tanaka-Montes}
	 T.~Tanaka and X.~Montes, Nucl. Phys. {\bf B582}, 259 (2000)
	 [hep-th/0001092]. 
 \bibitem{CGK}
	 C.~Csaki, M.~L.~Graesser and G.~D.~Kribs, Phys. Rev. {\bf D63},
	 065002 (2001) [hep-th/0008151]. 
 \bibitem{Mukohyama-Kofman}
	 S.~Mukohyama and L.~Kofman, Phys. Rev. {\bf D65}, 124025
	 (2002) [hep-th/0112115]. 
 \bibitem{Mukohyama-HDterm}
	 S.~Mukohyama, Phys. Rev. {\bf D65}, 084036 (2002)
	 [hep-th/0112205]. 
 \bibitem{Kudoh-Tanaka}
	 H.~Kudoh and T.~Tanaka, Phys. Rev. {\bf D65}, 104034 (2002)
	 [hep-th/0112013]; Phys. Rev. {\bf D67}, 044011 (2003)
	 [hep-th/0205041].
 \bibitem{CGR}
	 C.~Csaki, M.~Graesser and L.~Randall, Phys. Rev. {\bf D62},
	 045015 (2000) [hep-ph/9911406].
 \bibitem{KKOP}
	 P.~Kanti, I.~I.~Kogan, K.~A.~Olive and M.~Pospelov,
	 Phys. Lett. {\bf B468}, 31 (1999) [hep-ph/9909481]; P.~Kanti,
	 K.~A.~Olive and M.~Pospelov, Phys. Rev. {\bf D62}, 126004
	 (2000) [hep-ph/0005146]. 
 \bibitem{ADD}
	 N.~Arkani-Hamed, S.~Dimopoulos and G.~R.~Dvali,
	 Phys. Lett. {\bf B429}, 263 (1998) [hep-ph/9803315]; 
	 I.~Antoniadis, N.~Arkani-Hamed, S.~Dimopoulos and G.~R.~Dvali, 
	 Phys. Lett. {\bf B436}, 257 (1998) [hep-ph/9804398]; 
	 N.~Arkani-Hamed, S.~Dimopoulos and G.~R.~Dvali, Phys. Rev. 
	 {\bf D59}, 086004 (1999) [hep-ph/9807344]. 
 \bibitem{Carroll-Guica}
	 S.~M.~Carroll and M.~M.~Guica, ``Sidestepping the Cosmological
	 Constant with Football Shaped Extra Dimensions'',
	 hep-th/0302067. 
 \bibitem{Garriga-Porrati}
	 J.~Garriga and M.~Porrati, JHEP {\bf 0408}, 028 (2004)
	 [hep-th/0406158]. 
 \bibitem{Navarro}
	 I.~Navarro, Class. Quant. Grav. {\bf 20}, 3603 (2003)
	 [hep-th/0305014]; 
	 JCAP {\bf 0309}, 004 (2003) [hep-th/0302129]. 
 \bibitem{Vinet-Cline}
	 J.~Vinet and J.~M.~Cline,
	 Phys. Rev.{\bf D71}, 064011 (2005) [arXiv:hep-th/0501098];
	 Phys. Rev. {\bf D70}, 083514 (2004) [arXiv:hep-th/0406141].
 \bibitem{CDGV}
	 J.~M.~Cline, J.~Descheneau, M.~Giovannini and J.~Vinet,
	 JHEP {\bf 0306}, 048 (2003) [arXiv:hep-th/0304147].
 \bibitem{Israel}
	 W.~Israel, Nuovo Cim. {\bf B44}, 1 (1966); Erratum-ibid. 
	 {\bf B48}, 463 (1967). 
 \bibitem{Israel1975}
	 W.~Israel, Bull, Am. Phys. Soc. {\bf 20}, 98 (1975);
	 Phys. Rev. {\bf D15}, 935 (1977). 
 \bibitem{Geroch-Traschen}
	 R.~Geroch and J.~Traschen, Phys. Rev. {\bf D36}, 1017 (1987). 
 \bibitem{FIU}
	 V.~P.~Frolov, W.~Israel and W.~G.~Unruh, Phys. Rev. {\bf D39},
	 1084 (1989). 
 \bibitem{Stability}
	 Work in progress. Preliminary consideration suggests stability
	 at least near $\eta_{\pm}=1$ at low energy, but detailed
	 analysis remains to be done. 
 \bibitem{Lanczos}
	 C.~Lanczos, Phys. Zeits, {\bf 23}, 539 (1922); Ann. der
	 Phys. {\bf 74}, 518 (1924). 
 \bibitem{Shiromizu}
	 T.~Shiromizu, Y.~Himemoto and K.~Takahashi, Phys. Rev. 
	 {\bf D70}, 107303 (2004) [hep-th/0405071]; 
	 T.~Shiromizu, K.~Takahashi, Y.~Himemoto and S.~Yamamoto,
	 Phys.~Rev. {\bf D70}, 123524 (2004) [hep-th/0407268]. 
 \bibitem{Koyama-Koyama}
	 K.~Koyama and K.~Koyama, ``Gravitational backreaction of anti-D
	 branes in the warped compactification'', hep-th/0505256. 
\end{thebibliography}
\end{document}